
\input harvmac.tex
\input amssym.tex 
\overfullrule=0pt 


\def\a{\alpha}\def\d{\delta}
\def\g{\gamma}\def\k{\kappa}\def\m{\mu}\def\n{\nu}
\def\o{\omega}\def\t{\tau}
\def\L{\Lambda}\def\O{\Omega}
\def\CA{{\cal A}}
\def\CE{{\cal E}}\def\CF{{\cal F}}\def\CI{{\cal I}}
\def\CN{{\cal N}}
\def\IP{{\Bbb P}} 
\def\IR{{\Bbb R}} 
\def\IZ{{\Bbb Z}} 

\def\mbar{{\bar m}}
\def\nbar{{\bar n}}

\def\Re{\mathop{\rm Re}\nolimits}
\def\Im{\mathop{\rm Im}\nolimits}
\def\Vol{\mathop{\rm Vol}\nolimits}

\def\equals{\mathop{=}} 

\def\tilde{\widetilde}

\def\w{\wedge}

\def\half{{\textstyle{1\over2}}}

\def\smallfrac#1#2{{\textstyle{{#1}\over{#2}}}}

\def\dil{{\rm dil}}
\def\tdil{{\t_\dil}}
\def\fib{{\rm fib}}
\def\base{{\rm base}}

\def\flux{{\rm flux}}

\def\footlabel#1{\xdef#1{\the\ftno}} 
\def\nonnumberedsubsec#1{\ifnum\lastpenalty>9000\else\bigbreak\fi
\noindent{\it{#1}}\par\nobreak\medskip\nobreak}
\def\Sec#1{Sec.~{#1}}


\lref\AspKthree{
P.~S.~Aspinwall, ``K3 surfaces and string duality,''
arXiv:hep-th/9611137.
}

\lref\AspTASI{
P.~S.~Aspinwall, ``Compactification, geometry and duality: $\CN =
2$,'' arXiv:hep-th/0001001.
}

\lref\AspUbiquity{
P.~S.~Aspinwall and J.~Louis, ``On the Ubiquity of K3 Fibrations in
String Duality,'' Phys.\ Lett.\ B {\bf 369}, 233 (1996)
[arXiv:hep-th/9510234].
}

\lref\AspFluxes{
P.~S.~Aspinwall, `An analysis of fluxes by duality,''
arXiv:hep-th/0504036.
}

\lref\AspComment{
P.~Aspinwall, private communication.
}

\lref\Atiyah{
M.~F.~Atiyah and N.~J.~Hitchin, ``Low-Energy Scattering Of Nonabelian
Monopoles,'' Phys.\ Lett.\ A {\bf 107}, 21 (1985)\semi
M.~F.~Atiyah and N.~J.~Hitchin, ``Low-Energy Scattering Of Nonabelian
Magnetic Monopoles,'' Phil.\ Trans.\ Roy.\ Soc.\ Lond.\ A {\bf 315},
459 (1985).
}

\lref\BCOV{
M.~Bershadsky, S.~Cecotti, H.~Ooguri and C.~Vafa, ``Kodaira-Spencer
theory of gravity and exact results for quantum string amplitudes,''
Commun.\ Math.\ Phys.\ {\bf 165}, 311 (1994)
[arXiv:hep-th/9309140]\semi
M.~Bershadsky, S.~Cecotti, H.~Ooguri and C.~Vafa, ``Holomorphic
anomalies in topological field theories,'' Nucl.\ Phys.\ B {\bf 405},
279 (1993) [arXiv:hep-th/9302103].
}

\lref\TdualH{
P.~Bouwknegt, J.~Evslin and V.~Mathai, ``T-duality: Topology change
from $H$-flux,'' arXiv:hep-th/0306062;
S.~Fidanza, R.~Minasian and A.~Tomasiello, ``Mirror symmetric
$SU(3)$-structure manifolds with NS fluxes,'' arXiv:hep-th/0311122.
}

\lref\Buscher{
T.~H.~Buscher, ``A Symmetry Of The String Background Field
Equations,'' Phys.\ Lett.\ B {\bf 194}, 59 (1987)\semi
T.~H.~Buscher, ``Path Integral Derivation Of Quantum Duality In
Nonlinear Sigma Models,'' Phys.\ Lett.\ B {\bf 201}, 466 (1988).
}

\lref\ConifoldTrans{
P.~Candelas, P.~S.~Green and T.~H\"ubsch, ``Rolling Among Calabi-Yau
Vacua,'' Nucl.\ Phys.\ B {\bf 330}, 49 (1990)\semi
A.~Strominger, ``Massless black holes and conifolds in string
theory,'' Nucl.\ Phys.\ B {\bf 451}, 96 (1995) [arXiv:hep-th/9504090]\semi
B.~R.~Greene, D.~R.~Morrison and A.~Strominger, ``Black hole
condensation and the unification of string vacua,'' Nucl.\ Phys.\ B
{\bf 451}, 109 (1995) [arXiv:hep-th/9504145].
}

\lref\DasguptaMukhi{
K.~Dasgupta, D.~P.~Jatkar and S.~Mukhi, ``Gravitational couplings and
Z(2) orientifolds,'' Nucl.\ Phys.\ B {\bf 523}, 465 (1998)
[arXiv:hep-th/9707224].  
}

\lref\MITjunction{
O.~DeWolfe and B.~Zwiebach, ``String junctions for arbitrary Lie
algebra representations,'' Nucl.\ Phys.\ B {\bf 541}, 509 (1999)
[arXiv:hep-th/9804210].
}

\lref\DouglasGreene{
M.~R.~Douglas and B.~R.~Greene, ``Metrics on D-brane orbifolds,''
Adv.\ Theor.\ Math.\ Phys.\ {\bf 1}, 184 (1998)
[arXiv:hep-th/9707214].
}

\lref\FP{
A.~R.~Frey and J.~Polchinski, ``$\CN=3$ warped compactifications,''
Phys.\ Rev.\ D {\bf 65}, 126009 (2002) [arXiv:hep-th/0201029].
}

\lref\IntrinsicTorsion{
A.~Gray and L.~M.~Hervella, ``The sixteen classes of almost Hermitian
manifolds and their linear invariants,'' Ann.\ Math.\ Pura Appl.\ {\bf
123} 35 (1980)\semi
S.~Chiossi and S.~Salamon, ``The Intrinsic Torsion of $SU(3)$ and
$G_2$ Structures,'' in {\it Differential Geometry, Valencia 2001\/},
World Sci.\ Publishing, River Edge, NJ, 115 (2002)
[arXiv:math.DG/0202282]\semi
S.~Gurrieri, J.~Louis, A.~Micu and D.~Waldram, ``Mirror symmetry in
generalized Calabi-Yau compactifications,'' Nucl.\ Phys.\ B {\bf 654},
61 (2003) [arXiv:hep-th/0211102]\semi
G.~L.~Cardoso, G.~Curio, G.~Dall'Agata, D.~L\"ust, P.~Manousselis and
G.~Zoupanos, ``Non-K\"ahler string backgrounds and their five torsion
classes,'' Nucl.\ Phys.\ B {\bf 652}, 5 (2003) [arXiv:hep-th/0211118]\semi
J.~P.~Gauntlett, D.~Martelli and D.~Waldram, ``Superstrings with
intrinsic torsion,'' arXiv:hep-th/0302158.
}

\lref\GandH{
P.~Griffiths and J.~Harris, ``Principles of Algebraic Geometry,''
Wiley-Interscience, 1978.
}

\lref\Hawking{
S.~W.~Hawking,``Gravitational Instantons,'' Phys.\ Lett.\ A {\bf 60},
81 (1977).\semi
G.~W.~Gibbons and S.~W.~Hawking, ``Classification Of Gravitational
Instanton Symmetries,'' Commun.\ Math.\ Phys.\ {\bf 66}, 291 (1979).
}

\lref\Bubbles{
S.~Kachru, X.~Liu, M.~B.~Schulz and S.~P.~Trivedi, ``Supersymmetry
changing bubbles in string theory,'' JHEP {\bf 0305}, 014 (2003)
[arXiv:hep-th/0205108].
}

\lref\NewSUSY{
S.~Kachru, M.~B.~Schulz, P.~K.~Tripathy and S.~P.~Trivedi, ``New
supersymmetric string compactifications,'' JHEP {\bf 0303}, 061 (2003)
[arXiv:hep-th/0211182].
}

\lref\Moduli{
S.~Kachru, M.~B.~Schulz and S.~Trivedi, ``Moduli stabilization from
fluxes in a simple IIB orientifold,'' JHEP {\bf 0310}, 007 (2003)
[arXiv:hep-th/0201028].
}

\lref\KlebStrass{
I.~R.~Klebanov and M.~J.~Strassler, ``Supergravity and a confining
gauge theory: Duality cascades and $\chi$SB-resolution of naked
singularities,'' JHEP {\bf 0008}, 052 (2000) [arXiv:hep-th/0007191].
}

\lref\Kobayashi{
S.~Kobayashi, ``Differential Geometry of Complex Vector Bundles,''
Princeton University Press, 1987.
}

\lref\Kreuzer{
M.~Kreuzer and H.~Skarke,
``Reflexive polyhedra, weights and toric Calabi-Yau fibrations,''
Rev.\ Math.\ Phys.\  {\bf 14}, 343 (2002)
[arXiv:math.ag/0001106]\semi
M.~Kreuzer and H.~Skarke, ``Complete classification of reflexive
polyhedra in four dimensions,'' Adv.\ Theor.\ Math.\ Phys.\ {\bf 4},
1209 (2002) [arXiv:hep-th/0002240].
}

\lref\KreuzerWeb{
http://hep.itp.tuwien.ac.at/\~{}kreuzer/CY/
}

\lref\KreuzerPALP{
M.~Kreuzer and H.~Skarke, ``PALP: A Package for analyzing lattice
polytopes with applications to toric geometry,'' Comput.\ Phys.\
Commun.\ {\bf 157}, 87 (2004) [arXiv:math.na/0204356].
}

\lref\GrassiKlemm{
A.~Grassi and A.~Klemm, private communication.
}

\lref\McCleary{J.~McCleary, ``A Users Guide to Spectral Sequences,''
Cambridge University Press, 2000.}

\lref\OguisoCYQuotient{
K.~Oguiso and J.~Sakurai, ``Calabi-Yau threefolds of quotient 
type,'' math.AG/9909175.
}

\lref\OguisoThm{
K.~Oguiso, ``On Algebraic Fiber Space Structures on a Calabi-Yau
3-fold,'' Int.\ J.\ of Math.\ {\bf 4} 439 (1993).  }

\lref\TNHarmonic{
P.~J.~Ruback, ``The Motion Of Kaluza-Klein Monopoles,'' Commun.\
Math.\ Phys.\ {\bf 107}, 93 (1986).
}

\lref\Otorsion{
M.~B.~Schulz, ``Superstring orientifolds with torsion: O5 orientifolds
of torus fibrations and their massless spectra,'' Fortsch.\ Phys.\
{\bf 52}, 963 (2004) [arXiv:hep-th/0406001].
}

\lref\ProbeBrane{
N.~Seiberg,
``IR dynamics on branes and space-time geometry,''
Phys.\ Lett.\ B {\bf 384}, 81 (1996)
[arXiv:hep-th/9606017]\semi
N.~Seiberg and E.~Witten, ``Gauge dynamics and compactification to
three dimensions,'' arXiv:hep-th/9607163.
}

\lref\SW{
N.~Seiberg and E.~Witten, ``Electric-magnetic duality, monopole
condensation, and confinement in N=2 supersymmetric Yang-Mills
theory,'' Nucl.\ Phys.\ B {\bf 426}, 19 (1994) [Erratum-ibid.\ B {\bf
430}, 485 (1994)] [arXiv:hep-th/9407087]\semi
N.~Seiberg and E.~Witten, ``Monopoles, duality and chiral symmetry
breaking in N=2 supersymmetric QCD,'' Nucl.\ Phys.\ B {\bf 431}, 484
(1994) [arXiv:hep-th/9408099].
}

\lref\BprobeB{
A.~Sen, ``F-theory and Orientifolds,'' Nucl.\ Phys.\ B {\bf 475}, 562
(1996) [arXiv:hep-th/9605150]\semi
T.~Banks, M.~R.~Douglas and N.~Seiberg, ``Probing F-theory with
branes,'' Phys.\ Lett.\ B {\bf 387}, 278 (1996)
[arXiv:hep-th/9605199].
}

\lref\Sen{
A.~Sen, ``A note on enhanced gauge symmetries in M- and string
theory,'' JHEP {\bf 9709}, 001 (1997) [arXiv:hep-th/9707123].
}

\lref\SBW{
N.~I.~Shephard-Barron and P.~M.~H.~Wilson, ``Singular threefolds with
numerically trivial first and second chern class,'' J.\ Alg.\ Geom.\
{\bf 3} 265 (1994).
}

\lref\TripathyTrivedi{
P.~K.~Tripathy and S.~P.~Trivedi, ``Compactification with flux on $K3$
and tori,'' JHEP {\bf 0303}, 028 (2003) [arXiv:hep-th/0301139]\semi
C.~Angelantonj, R.~D'Auria, S.~Ferrara and M.~Trigiante, ``$K3\times
T^2/\IZ_2$ orientifolds with fluxes, open string moduli and critical
points,'' Phys.\ Lett.\ B {\bf 583}, 331 (2004)
[arXiv:hep-th/0312019].
}

\lref\Verlinde{
H.~Verlinde, ``Holography and compactification,'' Nucl.\ Phys.\ B {\bf
580}, 264 (2000) [arXiv:hep-th/9906182]\semi
C.~S.~Chan, P.~L.~Paul and H.~Verlinde, ``A note on warped string
compactification,'' Nucl.\ Phys.\ B {\bf 581}, 156 (2000)
[arXiv:hep-th/0003236].
}

\lref\ModStab{
J.~Polchinski and A.~Strominger, ``New Vacua for Type II String
Theory,'' Phys.\ Lett.\ B {\bf 388}, 736 (1996)
[arXiv:hep-th/9510227]\semi
K.~Becker and M.~Becker, ``M-Theory on Eight-Manifolds,'' Nucl.\
Phys.\ B {\bf 477}, 155 (1996) [arXiv:hep-th/9605053]\semi
S.~Gukov, C.~Vafa and E.~Witten, ``CFT's from Calabi-Yau four-folds,''
Nucl.\ Phys.\ B {\bf 584}, 69 (2000) [Erratum-ibid.\ B {\bf 608}, 477
(2001)] [arXiv:hep-th/9906070]\semi
K.~Dasgupta, G.~Rajesh and S.~Sethi, ``M theory, orientifolds and
$G$-flux,'' JHEP {\bf 9908}, 023 (1999) [arXiv:hep-th/9908088]\semi
S.~Sethi, ``Warped Compactifications,'' talk at ITP Conference on New 
Dimensions in String Theory and Field Theory,
http://online.kitp.ucsb.edu/online/susy\_c99/sethi/\semi
T.~R.~Taylor and C.~Vafa, ``RR flux on Calabi-Yau and partial
supersymmetry breaking,'' Phys.\ Lett.\ B {\bf 474}, 130 (2000)
[arXiv:hep-th/9912152]\semi
P.~Mayr, ``On supersymmetry breaking in string theory and its
realization in brane worlds,'' Nucl.\ Phys.\ B {\bf 593}, 99 (2001)
[arXiv:hep-th/0003198]\semi
B.~R.~Greene, K.~Schalm and G.~Shiu, ``Warped compactifications in M
and F theory,'' Nucl.\ Phys.\ B {\bf 584}, 480 (2000)
[arXiv:hep-th/0004103]\semi
G.~Curio, A.~Klemm, D.~L\"ust and S.~Theisen, ``On the vacuum structure
of type II string compactifications on Calabi-Yau spaces with
$H$-fluxes,'' Nucl.\ Phys.\ B {\bf 609}, 3 (2001)
[arXiv:hep-th/0012213]\semi
S.~B.~Giddings, S.~Kachru and J.~Polchinski, ``Hierarchies from fluxes
in string compactifications,'' Phys.\ Rev.\ D {\bf 66}, 106006 (2002)
[arXiv:hep-th/0105097]\semi
G.~Curio, A.~Klemm, B.~K\"ors and D.~L\"ust, ``Fluxes in heterotic and
type II string compactifications,'' Nucl.\ Phys.\ B {\bf 620}, 237
(2002) [arXiv:hep-th/0106155]\semi
G.~Dall'Agata, ``Type IIB supergravity compactified on a Calabi-Yau
manifold with H-fluxes,'' JHEP {\bf 0111}, 005 (2001)
[arXiv:hep-th/0107264]\semi
G.~Curio, B.~K\"ors and D.~L\"ust, ``Fluxes and branes in type II vacua
and M-theory geometry with $G_2$ and Spin(7) holonomy,'' Nucl.\ Phys.\
B {\bf 636}, 197 (2002) [arXiv:hep-th/0111165]\semi
J.~Louis and A.~Micu, ``Type II theories compactified on Calabi-Yau
threefolds in the presence of background fluxes,'' Nucl.\ Phys.\ B
{\bf 635}, 395 (2002) [arXiv:hep-th/0202168]\semi
O.~DeWolfe and S.~B.~Giddings, ``Scales and hierarchies in warped
compactifications and brane worlds,'' Phys.\ Rev.\ D {\bf 67}, 066008
(2003) [arXiv:hep-th/0208123]\semi
A.~R.~Frey and A.~Mazumdar, ``3-form induced potentials, dilaton
stabilization, and running moduli,'' Phys.\ Rev.\ D {\bf 67}, 046006
(2003) [arXiv:hep-th/0210254]\semi
M.~Berg, M.~Haack and B.~K\"ors, ``An orientifold with fluxes and
branes via T-duality,'' Nucl.\ Phys.\ B {\bf 669}, 3 (2003)
[arXiv:hep-th/0305183]\semi
S.~P.~de Alwis, ``On potentials from fluxes,'' Phys.\ Rev.\ D {\bf
68}, 126001 (2003) [arXiv:hep-th/0307084]\semi
A.~R.~Frey, ``Warped strings: Self-dual flux and contemporary
compactifications,'' arXiv:hep-th/0308156\semi
A.~Giryavets, S.~Kachru, P.~K.~Tripathy and S.~P.~Trivedi, ``Flux
compactifications on Calabi-Yau threefolds,'' JHEP {\bf 0404}, 003
(2004) [arXiv:hep-th/0312104]\semi
E.~Silverstein, ``TASI/PiTP/ISS lectures on moduli and
microphysics,'' arXiv:hep-th/0405068.
}

\lref\KaehlerStab{
E.~Witten, ``Non-Perturbative Superpotentials In String Theory,''
Nucl.\ Phys.\ B {\bf 474}, 343 (1996) [arXiv:hep-th/9604030]\semi
K.~Becker, M.~Becker, M.~Haack and J.~Louis, ``Supersymmetry breaking
and $\alpha'$-corrections to flux induced potentials,'' JHEP {\bf
0206}, 060 (2002) [arXiv:hep-th/0204254].
}

\lref\InflationdS{
S.~Kachru, R.~Kallosh, A.~Linde and S.~P.~Trivedi, ``De Sitter vacua
in string theory,'' Phys.\ Rev.\ D {\bf 68}, 046005 (2003)
[arXiv:hep-th/0301240]\semi
M.~Fabinger and E.~Silverstein, ``D-Sitter space: Causal structure,
thermodynamics, and entropy,'' arXiv:hep-th/0304220\semi
A.~R.~Frey, M.~Lippert and B.~Williams, ``The fall of stringy de
Sitter,'' Phys.\ Rev.\ D {\bf 68}, 046008 (2003)
[arXiv:hep-th/0305018]\semi
S.~Kachru, R.~Kallosh, A.~Linde, J.~Maldacena, L.~McAllister and
S.~P.~Trivedi, ``Towards inflation in string theory,'' JCAP {\bf
0310}, 013 (2003) [arXiv:hep-th/0308055]\semi
C.~P.~Burgess, R.~Kallosh and F.~Quevedo, ``de Sitter string vacua
from supersymmetric $D$-terms,'' JHEP {\bf 0310}, 056 (2003)
[arXiv:hep-th/0309187]\semi
E.~Silverstein and D.~Tong, ``Scalar speed limits and cosmology:
Acceleration from D-cceleration,'' arXiv:hep-th/0310221\semi
J.~P.~Hsu, R.~Kallosh and S.~Prokushkin, ``On brane inflation with
volume stabilization,'' JCAP {\bf 0312}, 009 (2003)
[arXiv:hep-th/0311077]\semi
A.~Buchel and R.~Roiban, ``Inflation in warped geometries,''
arXiv:hep-th/0311154\semi
F.~Koyama, Y.~Tachikawa and T.~Watari, ``Supergravity analysis of
hybrid inflation model from D3-D7 system,'' Phys.\ Rev.\ D {\bf 69},
106001 (2004) [arXiv:hep-th/0311191]\semi
H.~Firouzjahi and S.~H.~H.~Tye, ``Closer towards inflation in string
theory,'' Phys.\ Lett.\ B {\bf 584}, 147 (2004)
[arXiv:hep-th/0312020]\semi
A.~Buchel, ``On effective action of string theory flux
compactifications,'' Phys.\ Rev.\ D {\bf 69}, 106004 (2004)
[arXiv:hep-th/0312076]\semi
R.~Brustein and S.~P.~de Alwis, ``Moduli potentials in string
compactifications with fluxes: Mapping the discretuum,''
arXiv:hep-th/0402088\semi
A.~Saltman and E.~Silverstein, ``The scaling of the no-scale potential
and de Sitter model building,'' arXiv:hep-th/0402135\semi
L.~Kofman, A.~Linde, X.~Liu, A.~Maloney, L.~McAllister and
E.~Silverstein, ``Beauty is attractive: Moduli trapping at enhanced
symmetry points,'' arXiv:hep-th/0403001\semi
O.~DeWolfe, S.~Kachru and H.~Verlinde, ``The giant inflaton,''
arXiv:hep-th/0403123\semi
N.~Iizuka and S.~P.~Trivedi, ``An inflationary model in string
theory,'' arXiv:hep-th/0403203\semi
M.~Alishahiha, E.~Silverstein and D.~Tong, ``DBI in the sky,''
arXiv:hep-th/0404084\semi
M.~Berg, M.~Haack and B.~K\"ors, ``Loop corrections to volume moduli
and inflation in string theory,'' arXiv:hep-th/0404087\semi
A.~Buchel and A.~Ghodsi, ``Braneworld inflation,''
arXiv:hep-th/0404151\semi
F.~Denef, M.~R.~Douglas and B.~Florea, ``Building a better
racetrack,'' arXiv:hep-th/0404257\semi
V.~Balasubramanian, ``Accelerating universes and string theory,''
Class.\ Quant.\ Grav.\ {\bf 21}, S1337 (2004)
[arXiv:hep-th/0404075]\semi
J.~J.~Blanco-Pillado {\it et al.}, ``Racetrack inflation,'' JHEP {\bf
0411}, 063 (2004) [arXiv:hep-th/0406230]\semi
L.~Gorlich, S.~Kachru, P.~K.~Tripathy and S.~P.~Trivedi, ``Gaugino
condensation and nonperturbative superpotentials in flux
compactifications,'' arXiv:hep-th/0407130\semi
V.~Balasubramanian and P.~Berglund, ``Stringy corrections to K\"ahler
potentials, SUSY breaking, and the cosmological constant problem,''
JHEP {\bf 0411}, 085 (2004) [arXiv:hep-th/0408054]\semi
R.~Kallosh and A.~Linde, ``Landscape, the scale of SUSY breaking, and
inflation,'' arXiv:hep-th/0411011\semi
A.~Saltman and E.~Silverstein,
``A new handle on de Sitter compactifications,''
arXiv:hep-th/0411271.
}

\lref\Landscape{
L.~Susskind, ``The anthropic landscape of string theory,''
arXiv:hep-th/0302219\semi
M.~R.~Douglas, ``The statistics of string/M theory vacua,'' JHEP
{\bf 0305}, 046 (2003) [arXiv:hep-th/0303194]\semi
S.~Ashok and M.~R.~Douglas, ``Counting flux vacua,''
arXiv:hep-th/0307049\semi
T.~Banks, M.~Dine and E.~Gorbatov, ``Is there a string theory
landscape?,'' arXiv:hep-th/0309170\semi
M.~R.~Douglas, ``Statistics of string vacua,''
arXiv:hep-ph/0401004\semi
M.~Dine, ``Is there a string theory landscape: Some cautionary
notes,'' arXiv:hep-th/0402101\semi
M.~R.~Douglas, B.~Shiffman and S.~Zelditch, ``Critical points and
supersymmetric vacua,'' arXiv:math.cv/0402326\semi
F.~Denef and M.~R.~Douglas, ``Distributions of flux vacua,''
arXiv:hep-th/0404116\semi
A.~Giryavets, S.~Kachru and P.~K.~Tripathy, ``On the taxonomy of flux
vacua,'' arXiv:hep-th/0404243\semi
D.~Robbins and S.~Sethi, ``A barren landscape,''
arXiv:hep-th/0405011\semi
L.~Susskind, ``Supersymmetry breaking in the anthropic landscape,''
arXiv:hep-th/0405189\semi
M.~R.~Douglas, ``Statistical analysis of the supersymmetry breaking
scale,'' arXiv:hep-th/0405279\semi
B.~Freivogel and L.~Susskind, ``A framework for the landscape,''
arXiv:hep-th/0408133\semi
J.~P.~Conlon and F.~Quevedo, ``On the explicit construction and
statistics of Calabi-Yau flux vacua,'' JHEP {\bf 0410}, 039 (2004)
[arXiv:hep-th/0409215]\semi
J.~Kumar and J.~D.~Wells, ``Landscape cartography: A coarse survey of
gauge group rank and stabilization of the proton,'' Phys.\ Rev.\ D
{\bf 71}, 026009 (2005) [arXiv:hep-th/0409218]\semi
O.~DeWolfe, A.~Giryavets, S.~Kachru and W.~Taylor, ``Enumerating flux
vacua with enhanced symmetries,'' arXiv:hep-th/0411061\semi
R.~Blumenhagen, F.~Gmeiner, G.~Honecker, D.~L\"ust and T.~Weigand, ``The
statistics of supersymmetric D-brane models,''
arXiv:hep-th/0411173\semi
F.~Denef and M.~R.~Douglas, ``Distributions of nonsupersymmetric flux
vacua,'' arXiv:hep-th/0411183\semi
T.~Banks, ``Landskepticism or why effective potentials don't count
string models,'' arXiv:hep-th/0412129.
}

\lref\SoftMass{
R.~Blumenhagen, D.~L\"ust and T.~R.~Taylor, ``Moduli stabilization in
chiral type IIB orientifold models with fluxes,'' Nucl.\ Phys.\ B {\bf
663}, 319 (2003) [arXiv:hep-th/0303016]\semi
P.~G.~C\'amara, L.~E.~Ib\'a\~nez and A.~M.~Uranga, ``Flux-induced
SUSY-breaking soft terms,'' arXiv:hep-th/0311241\semi
J.~F.~G.~Cascales and A.~M.~Uranga, ``Chiral 4d string vacua with
D-branes and moduli stabilization,'' arXiv:hep-th/0311250\semi
J.~F.~G.~Cascales, M.~P.~Garc\'\i a del Moral, F.~Quevedo and
A.~M.~Uranga, ``Realistic D-brane models on warped throats: Fluxes,
hierarchies and moduli stabilization,'' JHEP {\bf 0402}, 031 (2004)
[arXiv:hep-th/0312051]\semi
M.~Gra\~na, T.~W.~Grimm, H.~Jockers and J.~Louis, ``Soft supersymmetry
breaking in Calabi-Yau orientifolds with D-branes and fluxes,''
arXiv:hep-th/0312232\semi
A.~Lawrence and J.~McGreevy, ``Local string models of soft
supersymmetry breaking,'' arXiv:hep-th/0401034\semi
T.~W.~Grimm and J.~Louis, ``The effective action of $\CN=1$ Calabi-Yau
orientifolds,'' arXiv:hep-th/0403067\semi
D.~L\"ust, S.~Reffert and S.~Stieberger, ``Flux-induced Soft
Supersymmetry Breaking in Chiral Type IIB Orientifolds with
D3/D7-Branes,'' arXiv:hep-th/0406092\semi
L.~E.~Ib\'a\~nez, ``The fluxed MSSM,'' arXiv:hep-ph/0408064\semi
P.~G.~C\'amara, L.~E.~Ib\'a\~nez and A.~M.~Uranga, ``Flux-induced
SUSY-breaking soft terms on D7-D3 brane systems,''
arXiv:hep-th/0408036\semi
F.~Marchesano and G.~Shiu, ``MSSM vacua from flux compactifications,''
arXiv:hep-th/0408059\semi
M.~Cveti\v c and T.~Liu, ``Supersymmetric Standard Models, Flux
Compactification and Moduli Stabilization,'' arXiv:hep-th/0409032\semi
F.~Marchesano and G.~Shiu, ``Building MSSM flux vacua,'' JHEP {\bf
0411}, 041 (2004) [arXiv:hep-th/0409132]\semi
D.~L\"ust, S.~Reffert and S.~Stieberger, ``MSSM with soft SUSY
breaking terms from D7-branes with fluxes,'' arXiv:hep-th/0410074\semi
F.~Marchesano, G.~Shiu and L.~T.~Wang, ``Model building and
phenomenology of flux-induced supersymmetry breaking on D3-branes,''
arXiv:hep-th/0411080\semi
A.~Font and L.~E.~Ib\'a\~nez, ``SUSY-breaking soft terms in a MSSM
magnetized D7-brane model,'' arXiv:hep-th/0412150.
}

\lref\SuperHiggs{
R.~D'Auria, S.~Ferrara and S.~Vaula, ``$\CN=4$ gauged supergravity and
a IIB orientifold with fluxes,'' New J.\ Phys.\ {\bf 4}, 71 (2002)
[arXiv:hep-th/0206241]\semi
R.~D'Auria, S.~Ferrara, F.~Gargiulo, M.~Trigiante and S.~Vaula,
``$\CN=4$ supergravity Lagrangian for type IIB on $T^6/\IZ_2$ in
presence of fluxes and D3-branes,'' JHEP {\bf 0306}, 045 (2003)
[arXiv:hep-th/0303049].
}


\Title{\vbox{\hbox{hep-th/0412270}\hbox{CALT-68-2533}}}
{\vbox{\baselineskip=22pt
\centerline{Calabi-Yau Duals of Torus Orientifolds}}}
\centerline{Michael B.~Schulz\footnote{$^\dagger$}{mschulz at
theory.caltech.edu}}
\bigskip\centerline{\it {California Institute of Technology 452-48}}
\centerline{{\it Pasadena, CA 91125 USA}}

\vskip .3in

We study a duality that relates the $T^6/\IZ_2$ orientifold with
$\CN=2$ flux to standard fluxless Calabi-Yau compactifications of type
IIA string theory.  Using the duality map, we show that the Calabi-Yau
manifolds that arise are abelian surface ($T^4$) fibrations over
$\IP^1$.  We compute a variety of properties of these threefolds,
including Hodge numbers, intersection numbers, discrete isometries,
and $H_1(X,\IZ)$.  In addition, we show that S-duality in the
orientifold description becomes T-duality of the abelian surface
fibers in the dual Calabi-Yau description.  The analysis is
facilitated by the existence of an explicit Calabi-Yau metric on an
open subset of the geometry that becomes an arbitrarily good
approximation to the actual metric (at most points) in the limit that
the fiber is much smaller than the base.

\Date{22 December 2004}


\newsec{Introduction}

In this investigation, we study a duality that relates the simplest
$\CN=2$ warped compactifications to standard type IIA Calabi-Yau
compactifications with no flux.  We view this investigation as a first
step toward the larger goal of understanding which warped
compactifications represent new string vacua, and which are just
alternative descriptions of more conventional fluxless
compactifications.

Warped compactifications, specifically those of D3/D7-type, are an
exciting arena in which we have begun to address several important
problems in string theory and its applications to cosmology and
particle physics.  One of the insights of this class of
compactifications is in understanding moduli stabilization.  Turning
on internal Neveu-Schwarz (NS) and Ramond-Ramond (RR) 3-form flux
generates a perturbative superpotential that generically stabilizes
all complex structure moduli and the
axion-dilaton~\refs{\ModStab,\FP,\Moduli,\TripathyTrivedi,\SuperHiggs}.
By combining this observation with mechanisms for K\"ahler moduli
stabilization and supersymmetry breaking, this class has been used to
construct metastable de~Sitter vacua and a variety realizations of
inflation in string theory~\refs{\KaehlerStab,\InflationdS}.  It is
also the ensemble within which statements have been made about the
``landscape'' of string theory vacua~\Landscape.  Most recently,
D3/D7-type warped compactifications have been used in MSSM-like model
building, where they provide an underlying theory of soft masses that
shows promise for explaining some of the usual assumptions on soft
masses and the $\mu$-term needed to avoid contradiction with
experiment~\SoftMass.  Finally, warped compactifications come with a
holographic interpretation (due to the warping) that often lets us
view field theory phenomena in an intuitive geometrical way~\Verlinde.
For example, the scale of gaugino condensation in the model of
Klebanov and Strassler is related to the size of a minimal 3-sphere in
the warped deformed conifold~\KlebStrass.

These D3/D7-type warped compactifications, with all of their desirable
features, are often presented in a way that suggests they are simply a
convenient alternative description of more conventional heterotic or
type II compactifications rather than independent vacua.  That is,
they make certain phenomena manifest that are also present in more
complicated ways in conventional type II or heterotic duals.  However,
it is not clear in which cases there actually exist conventional
duals.

As usual, the question is easiest to answer in the case of $\CN=4$
supersymmetry.  In this case, the only warped compactification with D3
branes is the $T^6/\IZ_2$ orientifold with 16 D3 branes and 64 O3
planes.  This background is dual to type I on $T^6$ (via six
\hbox{T-dualities}), and then to the heterotic string on $T^6$ (via
S-duality), and to type IIA on $\hbox{K3}\times T^2$ (via
heterotic/IIA duality).  The $\CN=1$ case is more difficult and we
will have nothing to say about it here.  On the other hand, the
$\CN=2$ case is nontrivial enough to be interesting, yet simple enough
to be tractable via standard string dualities.  We will focus on the
$\CN=2$ case in this investigation.

For $\CN=2$, there exist large web(s) of connected Calabi-Yau vacua.
One of the goals in asking about Calabi-Yau dual descriptions of
$\CN=2$ orientifolds is to improve our understanding of the
connectedness of string theory vacua---a likely prerequisite for
understanding potential mechanisms for vacuum selection in string
theory.  The strongest notion of vacuum connectedness is connectedness
in moduli space.  In this sense, all Calabi-Yau manifolds that are
hypersurfaces in 4D toric varieties are connected~\Kreuzer, and there
are expected to be other large webs of Calabi-Yau vacua that are
connected in the same way through flops and conifold
transitions~\ConifoldTrans.  Weaker notions of connectedness also
exist and could be relevant in cosmology, which probes more than just
the minima of potentials.  For example, all $T^6/\IZ_2$ orientifold
vacua have been shown to be connected through time-dependent bubbles,
including pairs of vacua that preserve unequal amounts of
supersymmetry~\Bubbles.  By relating Calabi-Yau vacua to $\CN=2$ vacua
of the $T^6/\IZ_2$ orientifold, we open the possibility of unifying
into one large class the two types of connected vacua just described.

An outline of the paper is as follows.  In \Sec{2} we review the
$T^6/\IZ_2$ orientifold~\refs{\FP,\Moduli}\ with $\CN=2$ flux.  The
space of warped compactifications with compact internal manifold that
preserve $\CN=2$ supersymmetry includes the $T^6/\IZ_2$ orientifold
and the $T^2/\IZ_2\times\hbox{K3}$ orientifold~\TripathyTrivedi.  We
restrict to $T^6/\IZ_2$ for simplicity.  After introducing the
necessary background on $T^6/\IZ_2$ in \Sec{2.1}, we derive the metric
moduli constraints resulting from $\CN=2$ flux in \Sec{2.2}, and the
complete massless spectrum in \Sec{2.3}.

In \Sec{3}, we describe the chain of dualities that we will use to
relate $T^6/\IZ_2$ with $\CN=2$ flux to a purely geometrical type IIA
Calabi-Yau compactification---three T-dualites followed by a 9-10
circle swap.  This duality was first sketched in Ref.~\NewSUSY. The
remainder of \Sec{3} is devoted to implementing the T-dualities.  In
\Sec{3.1}, we review the action of T-duality on NS flux.  Then, in
\Sec{3.2}, we perform the desired three T-dualities to obtain the
O6/D6 backround dual to $T^6/\IZ_2$.

In \Sec{4}, we perform a lift to M~theory on the $x^{10}$ circle and a
compactification to IIA on the $x^{9}$ circle.  We begin in \Sec{4.1}
with a qualitative overview of how the chain of dualities results in a
Calabi-Yau compactification.  Approximate metrics for the dual
Calabi-Yau manifolds are derived from the duality chain in \Sec{4.2}.
In \Sec{4.3}, we obtain the K\"ahler form and $(3,0)$-form in these
approximate metrics and show that they are closed.  \Sec{4.4} provides
a further discussion of the complex structure moduli and an
enumeration of the K\"ahler and complex structure moduli in the
approximate description.

The purpose of \Sec{5} is to establish an intuition for the sense in
which the leading order approximate M~theory lift encodes most of the
topological data that one would ever want to extract from the complete
lift.  This section is largely a review of Sen's treatment in
Ref.~\Sen.  We first explain why the leading order lift is only an
approximation to the exact lift whenever orientifold planes are
present in the IIA background.  We then present examples of increasing
complexity, explaining how configurations of D6 branes and O6 planes
relate to the homology cycles and harmonic forms in the M~theory lift.

We get to the heart of the paper in \Sec{6}.  Using the leading order
description of the dual Calabi-Yau manifolds, we compute a basis of
harmonic representatives of $H^2(X_6,\IZ)$ in \Sec{6.1}.  We then use
this basis to compute the intersection numbers.  In \Sec{6.2}, we
demonstrate that these Calabi-Yau manifolds are abelian surface
fibrations over $\IP^1$.  Having identified the fibration structure,
we discuss the homology duals to the cohomology basis in \Sec{6.3}.
\Sec{6.4} contains a partially heuristic check of the fibration
result using a theorem by Oguiso and a feature of the topological
amplitude $F_1$ that we deduce from the $T^6/\IZ_2$ dual description.

In Secs.~7 and 8, we further sharpen our understanding of the
Calabi-Yau geometry.  In \Sec{7}, we identify discrete gauge
symmetries of the $T^6/\IZ_2$ orientifold that arise from $U(1)$ gauge
symmetries that are incompletely Higgsed by non-minimally-coupled
scalars.  These gauge symmetries tell us about torsion cycles and
discrete isometries in the dual Calabi-Yau manifolds.  In \Sec{8}, we
identify S-duality of the $T^6/\IZ_2$ orientifold with T-duality of
the abelian surface fibers in the Calabi-Yau dual description.

Finally, in \Sec{9}, we report attempts to identify some of the dual
Calabi-Yau manifolds in terms of known constructions.  From a duality
based computation of the second Chern class of these Calabi-Yau
manifolds, we rule out the possibility that they are quotients of
$T^6$.  This implies one of two possible interesting results, one
mathematical and one physical, depending upon whether the Calabi-Yau
duals contain rational curves.  If they do, then we infer the
existence of nonperturbative corrections in the $T^6/\IZ_2$
orientifold that have not been previously deduced by other means.  If
they do not, then we have mathematically interesting examples of
Calabi-Yau manifolds other than quotients of abelian threefolds that
do not contain rational curves.

In \Sec{10}, we conclude.


\newsec{The $T^6/\IZ_2$ orientifold with $\CN=2$ flux}

\subsec{Review of $T^6/\IZ_2$}

The $T^6/\IZ_2$ orientifold is defined by compactifying type IIB
string theory on $T^6$ and then quotienting by the $\IZ_2$ orientifold
operation
\eqn\ZtwoOrientifold{\IZ_2\colon\quad\Omega(-1)^{F_L}\CI_6.}
Here $\Omega$ is worldsheet orientation reversal, $(-1)^{F_L}$ is
left-moving fermion parity,\foot{The second factor is need to ensure a
supersymmetric spectrum.  For light states, i.e., those that
oscillatorwise are massless, $(-1)^{F_L}$ acts as $-1$ on left-moving
Ramond sector states and $+1$ on left-moving Neveu-Schwarz sector
states.}  and $\CI_6$ is inversion of the $T^6$. We take the
coordinates of the $T^6$ to have unit periodicity, $x^m\cong x^m+1$.
The inversion $\CI_6$ acts as $x^m\to -x^m$ on the $T^6$, with fixed
points at $x^m = 0,1/2$ for $m=1,\ldots,6$.  These are the locations
of 64 O3 planes.

For consistency of the model, a D3 charge cancellation condition on
the $T^6$ must be satisfied.  The condition is
\eqn\GaussLaw{2M + N_\flux = 32,}
where $M$ is the number of independent D3 branes and $N_\flux$ is
defined by
\eqn\Nflux{N_\flux = {1\over(2\pi)^4\a'^2}\int_{T^6} H_{(3)}\w F_{(3)}.}
This condition is the integral form of the $\tilde F_{(5)}$ Bianchi
identitity.  We assume that there is no localized flux on the O3
planes.  Then, on the $\IZ_2$ covering space $T^6$, there are $M$ D3
branes plus $M$ $\CI_6$ image branes, and the flux is quantized as
\eqn\FluxQuant{H_{(3)},F_{(3)}\in (2\pi)^2\a'H^3(T^6,2\IZ).}
The $2\IZ$ quantization condition on the covering space $T^6$ is a
$\IZ$ quantization condition on $T^6/\CI_6$.  It guarantees integer
periods of $F_{(3)}$ and $H_{(3)}$ over cycles in $T^6/\CI_6$ that
descend from half-cycles in $T^6$.

Finally, the geometry is warped.  The 10D string frame metric is
\eqn\StringMetric{ds^2 = Z^{-1/2}\eta_{\m\n}dx^\m dx^\n +
Z^{1/2}ds_{T^6}^2,}
where the warp factor satisfies the Poisson equation
\eqn\WarpEq{-\nabla^2_{T^6}Z = (2\pi)^4\a'^2g_s
\biggl(\sum_I Q_I{\d^6(x-x_I)\over\sqrt g_{T^6}} + {N_\flux\over
V_{T^6}}\biggr).}
The dilaton profile is constant in the internal directions,
\eqn\DilProfile{e^{\phi} = g_s.}
In Eq.~\WarpEq, the sum runs over $M$ D3 branes, $M$ image D3 branes,
and 64 O3 planes.  The charge $Q$ is normalized so that $Q_{\rm D3} =
1$ and $Q_{\rm O3} = -1/2$.

The simplest way to satisfy Eq.~\GaussLaw\ is with $M=16$ D3 branes
and no flux.  Then the model preserves $\CN=4$ supersymmetry and is
T-dual to type I on $T^6$, via T-duality in the six torus
directions.\foot{For each pair of directions, the T-duality takes
$\CI_{2p}$ to $\CI_{2p-2}$ and introduces another factor of
$(-1)^{F_L}$ in Eq.~\ZtwoOrientifold.}  More generally, we can trade
off some or all of the D3 branes for quantized flux while still
satisfying Eq.~\GaussLaw.  In the low energy effective field theory,
the fluxes are the charges of a 4D $\CN=4$ gauged supergravity theory.
In this theory, there is a superhiggs mechanism \SuperHiggs\ that
spontaneously breaks the supersymmetry to $\CN<4$.  In terms of the
torus length scale $R$, the superhiggs scale is $\a'/R^3$.  (The
factor of $\a'$ arises from the quantization condition \FluxQuant, and
the factor of $R^3$ from the volume of the 3-cycles carrying the
flux.)  For $R\gg 1/\sqrt{\a'}$, this scale is much lower than the
Kaluza-Klein scale $1/R$, which in turn is much lower than the string
scale.

\subsec{$\CN=2$ flux and constraints on metric moduli}

Up to $SL(2,\IZ)$ duality of the axion-dilaton and $SL(6,\IZ)$ change
of lattice basis of the $T^6$, the only known class of flux that
preserves $\CN=2$ supersymmetry is~\refs{\Moduli,\Otorsion}
\eqn\NtwoFlux{\eqalign{F_{(3)}/\bigl((2\pi)^2\a'\bigr)
& = 2m\bigl(dx^4\w dx^6 + dx^5\w dx^7\bigr)\w dx^9,\cr
H_{(3)}/\bigl((2\pi)^2\a'\bigr)
& = 2n\bigl(dx^4\w dx^6 + dx^5\w dx^7\bigr)\w dx^8.}}
Here $m$ and $n$ are positive integers satisfying the D3 charge
cancellation condition
\eqn\NtwoGauss{4mn + M = 16.}
Note that interchange of $m$ and $n$ has the interpretation of
S-duality followed by a $90^\circ$ rotation in the $89$-directions.
We will return to this $m\leftrightarrow n$ duality in \Sec{8}.

The superhiggs mechanism for this class of flux was discussed recently
in Ref.~\Otorsion.  The supersymmetry conditions require that $G_{(3)}
= F_{(3)} - \tdil H_{(3)}$ be primitive and of Hodge type $(2,1)$.
(Primitivity means that $J\w G_{(3)} = 0$, where $J$ is the K\"ahler
form.)  This constrains the torus to factorize as
\eqn\torusfact{T^6 = T^4_{\{4567\}}\times T^6_{\{89\}}}
with respect to both K\"ahler and complex structure, and then imposes
some additional constraints.  A convenient parametrization of the
metric moduli is obtained by writing the $T^2$ metric as
\eqn\TtwoMetric{ds^2_{T^2} = {v_3\over\Im\t_3}\bigl|dx^8+\t_3
dx^9\bigr|^2,}
and then writing the $T^4$ as a flat fibration of $T^2_{\{45\}}$ over
$T^2_{\{67\}}$:
\eqn\TtwoMetric{ds^4_{T^2} = {v_1\over\Im\t_1}\bigl|(dx^4+a^4) + \t_1
(dx^5 + a^5)|^2 + {v_2\over\Im\t_2}\bigl|dx^6+\t_2 dx^7\bigr|^2.}
Here the flat connections $a^4$ and $a^5$ are constant 1-forms on
$T^2_{\{67\}}$.  The $\Im\t_i$ factors in the denomenators are
necessary so that the $v_i$ are the volumes of the corresponding
2-tori.  In terms of this parameterization, the further constraints on
the metric moduli that follow from the supersymmetry conditions are
\eqn\OMetricModuli{\t_1\t_2 = -1,\quad (m/n)\tdil\t_3 = -1,\quad
(a^4)_7 = (a^5)_6,}
with $v_1,v_2,v_3$ arbitrary.

\subsec{The massless spectrum of $T^6/\IZ_2$ with $\CN=2$ flux}

The superhiggs mechanism involves more than just the metric moduli.
Before taking into account the masses due to the flux, the massless
bosonic fields preserved by the orientifold projection are
\eqna\ApproxModuli
$$\hbox{Bulk:}\quad
\left\{\matrix{\strut\hfill1&\hbox{4D graviton}\hfill & g_{\m\n},\hfill\cr
\strut\hfill12&\hbox{4D vectors}\hfill & b_{(2)m\m},\quad c_{(2)m\m},\hfill\cr
\strut\hfill38&\hbox{4D scalars}\hfill & c_{(4)mnpq},\quad g_{mn},\quad\tdil
= c_{(0)} + i e^{-\phi},\hfill}\right.\eqno\ApproxModuli a$$
from the closed string sector, and
$$\hbox{Branes:}\quad
M\times\left\{\matrix{\strut1 & \hbox{4D vector}\hfill &\hfill A^I{}_\m,\cr
\strut6 & \hbox{4D scalars}\hfill &\hfill \Phi^{I\,m},}\right.
\eqno\ApproxModuli b$$
from the open string sector.  In 4D $\CN =4$ terms, these fields fill
out the bosonic content of one gravity multiplet and $6+M$ vector
multiplets.

Now let us take into account the superhiggs mechanism due to the flux.
The constraints \OMetricModuli\ leave a total of 10 real metric
moduli: $v_1,v_2,v_3,$ 2 independent $\t,$ and 3 independent
$(a^m)_n$.  The kinetic terms for the $c_{(4)}$ scalars are
proportional to
\eqn\AxionKinetic{\bigl|dc_{(4)} - F_{(3)}\w b_{(2)}
+ H_{(3)}\w c_{(2)}\bigr|^2,}
from the flux kinetic term $\bigl|\tilde F_{(5)}\bigr|^2$.  (See
Ref.~\FP\ for a detailed discussion.)  The $c_{(4)}$ scalars are
axionically coupled to the vectors $b_{(2)m\m}$ and $c_{(2)n\m}$ with
charges given by $H_{(3)}$ and $F_{(3)}$, respectively.  One finds
that 9 of the vectors eat 9 of the $c_{(4)}$ axions, leaving a total
of 3 massless vectors and 6 massless $c_{(4)}$ axions \Otorsion.  In
the open string sector, none of the massless D3 worldvolume fields is
lifted by the flux.  All together, these massless fields fill out the
bosonic content of one $\CN=2$ gravity multiplet, $2+M$ vector
multiplets, and $3+M$ hypermultiplets.


\newsec{The O6/D6 dual of $T^6/\IZ_2$}

We will ultimately relate the above $T^6/\IZ_2$ orientifold vacua to
standard type IIA Calabi-Yau vacua by performing three T-dualities
followed by a 9-10 circle swap (i.e., lifting to M theory on $x^{10}$
and then compactifying on $x^9$).\foot{The existence of this chain of
dualities was first discussed in Ref.~\NewSUSY\ following suggestions
by P.~Berglund and N.~Warner.}  However, before attacking the full
problem, let us first review the action of T-duality on NS flux in a
simplified context~\refs{\NewSUSY}.

\subsec{Warm-up: the action of T-duality on NS flux}

Consider a $T^3$ in the $T^6/\IZ_2$ orientifold, such that $H_{(3)}$
through this $T^3$ is nonzero.  For simplicity, we neglect the warp
factor and take the metric on this $T^3$ to be
$$ds^2 = dx^2+dy^2+dz^2.$$
(It will be easy to include the warp factor and nontrivial metric when
we discuss the case of interest in the next section.)  Let the NS flux
and potential be
$$H_{(3)} = Ndx\w dy\w dz,\quad B_{(2)} = Nxdy\w dz.$$
Now T-dualize in the $z$-direction.  From the Buscher rules~\Buscher,
the resulting metric and NS $B$-field are
$$ds^2 = dx^2 + dy^2 + (dz+Nxdy)^2,\quad B_{(2)}=H_{(3)}=0.$$ The
$T^3$ metric has been replaced by an $S^1$ fibration, or equivalently
$U(1)$ principal bundle, over $T^2$.  The $U(1)$ connection is $\CA =
Nxdy$ and the curvature is $\CF = d\CA = Ndx\w dy$.  The topology of
the fibration is characterized by the Chern class $\bigl[\CF\bigr]$.

In other words, T-duality has interchanged the following two
fibrations: (i) the explicit geometrical $S^1$ fibration of the
isometry direction, and (ii) the formal $\tilde S^1$ fibration of
connection $\tilde\CA = -\int_{S^1}B_{(2)}$ and curvature $\tilde\CF =
\int_{S^1}H_{(3)}$.  More generally, as long as there is an isometry that
allows us to perform a T-duality, we can define both fibrations.
Therefore, T-duality can always be interpretated as this sort of
interchange.  (See Refs.~\refs{\TdualH,\Otorsion}\ for recent
discussions.)

\subsec{$T^6/\IZ_2$ after three T-dualities}

Now let us return to the problem of interest.  Starting from the
$T^6/\IZ_2$ orientifold with $\CN=2$ flux, we can perform successive
T-dualities in the $x^4$, $x^5$, and $x^9$ directions.  The resulting
10D string frame metric is
\eqna\DualOMetric
$$ds^2 = Z'^{-1/2}\bigl(\eta_{\m\n}dx^\m dx^\n +
ds^2_{T^3\,\fib}\bigr) + Z'^{1/2}
ds^2_{T^3\,\base},\eqno\DualOMetric a$$
where
$$\eqalignno{\strut ds^2_{T^2\,\fib} &= {v'_1\over\Im\t'_1}
\bigl|\eta^4+\t'_1\eta^5\bigr|^2 + {R'_9}^2 {dx^9}^2,
&\DualOMetric b\cr
\strut ds^2_{T^2\,\base} &= {v_2\over\Im\t_2}
\bigl|dx^6+\t_2 dx^7\bigr|^2 + {R_8}^2 {dx^8}^2.
&\DualOMetric c}$$
Here, $\eta^m = dx^m+\CA^m$ for $m=4,5$, where the connections
have curvatures
\eqn\Curvatures{\CF^4 = d\CA^4 = 2n dx^6\w dx^8\quad\hbox{and}
\quad\CF^4 = d\CA^5 = 2n dx^7\w dx^8.}
Just as a single T-duality gave us a circle fibration in the previous
section, the three T-dualities give us an $(S^1)^3=T^3$ fibration.
However, since there were only two components of NS flux before the
T-dualities (cf. Eq.~\NtwoFlux), only two of the $S^1$ fibrations are
nontrivial.

The primed metric moduli and string coupling after the T-dualities are
related to unprimed $T^6/\IZ_2$ quantities by
\eqn\PrimedModuli{v'_1 = (2\pi)^4\a'^2/v_1,\quad R'_9 =
(2\pi)^2\a'/R_9,\quad \t'_1 = -1/\t_1,\quad g'_s = g_s 
(2\pi)^6\a'^3/(R_4R_5R_9).}
These moduli satisfy the new constraints
\eqn\PrimedConstraints{(n/m)v'_1 = g'_s R_8,\quad \t'_1 = \t_2.}
The warp factor $Z'$ is related to $Z$ by averaging over the T-duality
directions.  It satisfies the equation
\eqn\WarpEqPrime{-\nabla^2_{T^3\,\base}Z' = 2\pi\sqrt{\a'}g'_s
\biggl(\sum_I Q_I{\d^3(x-x_I)\over\sqrt g_{T^3\,\base}} + {N_\flux\over
V_{T^3\,\base}}\biggr),}
where now $Q_{\rm D6}=1$ and $Q_{\rm O6} = -4$.  Each T-duality adds
one dimension to the O~planes and D~branes.  After the three
T-dualities, there are $M$ D6 branes, $M$ image D6 branes, and 8 O6
planes filling 4D spacetime and wrapping the $T^3$ fiber.  The O6
planes are located at the the fixed points of $\CI_3$ on the
$T^3_\base$, where $\CI_3$ takes $(x^6,x^7,x^8)$ to $(-x^6,-x^7,-x^8)$.
The solution to Eq.~\WarpEqPrime\ can be formally expressed as
\eqn\ZprimeSoln{Z' = 1 + 2\pi\sqrt{\a'}g'_s\sum_I Q_I G(x,x_I),}
where the Green's function $G(x,x')$ for the Poisson equation on
$T^3_\base$ is defined by
\eqn\GreensFn{-\nabla^2_{T^3}G(x,x') = {\d^3(x-x')\over\sqrt
g_{T^3}}  - {1\over V_{T^3}},\quad\hbox{with}\quad 
\int_{T^3}d^3x\sqrt g_{T^3}G(x,x') = 0.}
The leading constant of unity in Eq.~\ZprimeSoln\ has been chosen to
ensure that the warp factor drops out of the metric in the limit
$g'_s\to0$.

In terms of $Z'$, the new dilaton profile is
\eqn\DilPrime{e^{(\phi'-\phi'_0)} = Z'^{-3/4},\quad\hbox{where}\quad
e^{\phi'_0} = g'_s.}
Finally, the only flux after the three T-dualities is
\eqn\twoformflux{F_{(2)} = g'^{-1}_s *_3dZ' -
2m(2\pi\sqrt{\a'})\bigl(\eta^4\w dx^7-\eta^5\w dx^6\bigr).}


\newsec{M theory lift and dual IIA Calabi-Yau compactification}

\subsec{Overview}

In the strong coupling limit, the type IIA orientifold of the previous
section lifts to a purely geometrical M theory compactification with
no flux.  Via the standard identifications, the $F_{(2)}$ flux
determines the fibration of the $x^{10}$ circle over the type IIA
geometry, and the dilaton determines the size of this circle at each
point.  The only potentially singular objects in the lift are the O6
planes and D6 branes.  However, as long as these objects are not
coincident, each D6 brane lifts to a geometry that is locally a smooth
Taub-NUT space times $\IR^{6,1}$, and each O6 plane to a geometry that
is locally a smooth Atiyah-Hitchin space times $\IR^{6,1}$.  Before
the lift, the $x^9$ direction is not fibered, and does not appear in
$F_{(2)}$.  As we will see, the warp factors also cancel in the right
way so that after the lift, neither the Minkowski metric nor the $x^9$
metric is mixed with the other directions through a multiplicative
warp factor.  Therefore, the M theory geometry factorizes as
$\IR^{3,1}\times S^1_{\{9\}}\times X_6$, where $X_6$ is smooth.  Since
this background preserves $\CN=2$ supersymmetry, $X_6$ is necessarily
a Calabi-Yau threefold.  Compactifying on $S^1_{\{9\}}$ gives a
standard Calabi-Yau compactification of type IIA string theory.  This
proves the desired result:
$$T^6/\IZ_2\hbox{ with $\CN=2$ flux}\quad\leftrightarrow\quad
\hbox{IIA on a Calabi-Yau }X_6.$$

The natural question that arises is Which Calabi-Yau threefold?  We
immediately know at least one piece of information.  We computed the
massless spectrum in the dual $T^6/\IZ_2$ orientifold, and found that
there were $M+2$ vector multiplets and $M+3$ hypermultiplets, where
$M$ was the number of D3 branes.  This tells us that the Hodge numbers
of $X_6$ are
\eqn\HodgeNos{h^{1,1} = h^{2,1} = M+2.}
From the D3 charge cancellation condition $M=16-4mn$, the possible
values of $M$ are $M=0,4,8,12$, with multiplicities that are further
distinguished by the choice of positive integers $m$ and $n$.

Beyond the Hodge numbers, we would like to compute more detailed
properties of the Calabi-Yau duals.  What are the intersection
numbers?  Is there a fibration structure?  Is $X_6$ simply-connected
or is there a nontrivial fundamental group?  A compact Calabi-Yau
cannot have continuous isometries, but discrete isometries are
allowed.  Does $X_6$ have any disrete isometries?  Finally, we
identified the interchange of $m$ and $n$ as S-duality in the
$T^6/\IZ_2$ orientifold.  What is the corresponding duality in the
Calabi-Yau description?

\subsec{Dual M theory and type IIA approximate metrics}

In our conventions, the leading order relation between the 11D
M~theory metric and the type IIA dilaton, 1-form potential, and string
frame metric is
\eqn\MmetricGeneral{ds^2_{11} = e^{-2(\phi-\phi_0)/3}ds^2_{\rm IIA} 
+ e^{4(\phi-\phi_0)/3}{R_y}^2\bigl(dy + A\bigr)^2,}
where $R_y = 2\pi\sqrt{\a'}g_s$ and $A = C_{(1)}/(2\pi\sqrt{\a'})$.
Consequently, the leading order M~theory metric from the lift of the
O6/D6 dual of $T^6/\IZ_2$ is
\eqn\Mmetric{ds^2_{11} = \eta_{\m\n}dx^m dx^n + ds^2_6 + {R'_9}^2
{dx^9}^2,}
where
\eqn\CYmetricM{ds^2_6 = Z'^{-1}{R_{10}}^2 \bigl(dx^{10}+A\bigr)^2
+ {v'_1\over\Im\t'_1}\bigl|\eta^4+\t'_1\eta^5\bigr|^2 + Z'\Bigl(
{v_2\over\Im\t_2}\bigl|dx^6+\t_2 dx^7\bigr|^2 + {R_8}^2 {dx^8}^2
\Bigr),}
with coordinates identified modulo
\eqn\CYZtwo{(x^6,x^7,x^8,x^{10})\to-(x^6,x^7,x^8,x^{10}).} 
The 1-forms $\eta^4, \eta^5$ and corresponding circle fibrations are
defined after Eq.~\DualOMetric{}.  The fibration of the $x^{10}$
circle is defined by
\eqn\dA{dA = R_{10}^{-1}*_3dZ' - 2m\bigl(\eta^4\w dx^7 - \eta^5\w
dx^6\bigr),}
where $*_3$ is the Hodge star operator in the 3D base metric
\DualOMetric{c}\ in the 678-directions.  The 11D metric moduli satisfy
the relations
\eqn\ElevenDConstr{v'_1 = (m/n) R_8 R_{10},\quad \t'_1 = \t_2.}

From the M theory background, we can then compactify on $S^1_{\{9\}}$
to obtain the purely geometrical type IIA background
\eqn\IIACYcompact{ds^2_{\rm IIA} = \eta_{\m\n}dx^m dx^n + ds^2_6,}
with arbitrary string coupling.  In this dual IIA compactification,
the Regge slope and string coupling determine the M theory circle
radius $R'_9 = 2\pi\sqrt{\a'(\rm CY)}g_s(\rm CY)$.  The relation to
the 11D Planck scale is
\eqn\ElevenDPlanck{M_{11}^{-1} = g'^{1/3}_s\a'^{1/2} =
g_s^{1/3}(\rm CY)\a'^{1/2}({\rm CY}),}
where $1/(2\k_{11}^2) = (2\pi)^{-8}{M_{11}}^9$ is the prefactor
multiplying the 11D supergravity action.

\subsec{K\"ahler form and (3,0) form}

The metric \CYmetricM\ is {\it not} the Calabi-Yau metric on the
smooth Calabi-Yau $X_6$, but rather an approximation to it, in a sense
that we will make precise in the next section.  It is, however, a
Calabi-Yau metric on the open space $Z'>0$ for which the metric is
positive definite.  This is guaranteed by the fact that throughout the
classical supergravity dualities we have satisfied the equations for
an $\CN=2$ supersymmetric background.  As an independent check of this
claim, first note that there is a natural complex pairing of
coordinates in which the $(1,0)$-forms are
\eqn\ComplexPairing{\eqalign{\eta^{z^1} &= \eta^4+\t_2\eta^5,\cr
dz^2 &= dx^6+\t_2 dx^7\cr
\eta^{z^3} &= (dx^{10} + A) - it Z' dx^8,}}
where $t = R_8/R_{10}$.  For future convenience, let us define the
K\"ahler moduli $h$ and $s$ by
\eqn\hsDef{\bigl(v'_1, v_2, R_8R_{10}\bigl) = (\mbar h, 2s, \nbar h),
\quad\hbox{where}\quad (\mbar,\nbar) = (m,n)/\gcd(m,n).} 
These parameters will turn out to be K\"ahler moduli relative to basis
elements of integer cohomology.  The K\"ahler form and (3,0) form for
the complex structure \ComplexPairing\ and metric $ds^2_6$ are
\eqn\JO{\eqalign{J &= \mbar h \eta^4\w\eta^5 + 2sZ'dx^6\w dx^7
+ \nbar h dx^8\w(dx^{10} + A),\cr
\O &= \biggl({\mbar h\cdot2s\cdot\nbar h\over
\bigl(\Im\t_2\bigr)^2t}\biggr)^{1/2}\eta^{z^1}\w dz^2\w \eta^{z^3}.}}
Here, we have normalized $\O$ so that $\smallfrac{i}{8}\O\w\O =
\smallfrac{1}{6}J\w J\w J = \Vol_6.$  Using Eqs.~\Curvatures\ and \dA,
it is possible to show that $dJ = d\O = 0$.  So, the intrinsic
torsion~\IntrinsicTorsion\ vanishes and the region $Z'>0$ is
Calabi-Yau.

\subsec{K\"ahler and complex structure moduli}

From the superhiggs mechanism in the original $T^6/\IZ_2$ orientifold,
it is possible to show that the only massless deformation of $C_{(1)}$
in the dual O6/D6 orientifold is $\d C_{(1)}\propto dx^8$.  This
translates into a Calabi-Yau modulus $\delta A = a dx^8$.  Therefore,
let us write
\eqn\Aa{A = A_0 + adx^8,}
and define
\eqn\tauinverse{\t^{-1} = a - it.}
Then, Eq.~\ComplexPairing\ can be alternatively written as
\eqn\CpxPairing{\eqalign{\eta^{z^1} &= dx^4+\t_2 dx^5 + \CA^{z^1},\cr
dz^2 &= dx^6+\t_2 dx^7,\cr
\eta^{z^3} &= dx^{10}+\t^{-1}dx^8 + \CA^{z^3},}}
where
\eqn\CpxCurvatures{\eqalign{\CF^{z^1} &= d\CA^{z^1}=2ndz^2\w dx^8,\cr
\CF^{z^3} &= d\CA^{z^3}=\partial_8 \hat Z
dx^6\w dx^7- 2m\bigl(\eta^4\w dx^7 - \eta^5\w dx^6\bigr),}}
and $\hat Z$ is defined by
\eqn\Zhat{Z'= 1+\Bigl({\nbar h\over2s}\Bigr) \hat Z.}
The quantity $\hat Z$ satisfies the rescaled Poisson equation
\eqn\rescPoisson{\Bigl({\partial_8}^2 + {\nbar
ht\over2s}\nabla^2_{\hat T^2}\Bigr) 
\hat Z = \sum_I Q_I\bigl(\d^3(x-x_I)-1\bigr),\qquad
\int\hat Z d^3x = 0,}
where $\hat T^2$ is the torus obtained from $T^2_{\{6,7\}}$ by
rescaling to unit area.  In the limit of small relative K\"ahler
modulus $h/s\ll1$, the warp factor $Z'$ is positive and the metric
\CYmetricM\ is positive definite everywhere except in a small
neighborhood of the $\CI_3$ fixed loci.

The complete list of K\"ahler and complex structure moduli is
$$\vbox{\settabs\+&Complex structure Moduli &\qquad $\t$, $\t_2$, and
$M$ complex dof in $\hat Z$&\qquad ($2+M$ total)\cr
\+&\hfill K\"ahler moduli:&\qquad $h$, $s$, and $M$ real dof in $\hat
Z$\hfill& \qquad ($2+M$ total),\hfill\cr
\vskip5pt
\+&\hfill Complex structure moduli:&\qquad $\t$, $\t_2$, and $M$
complex dof in $\hat Z$\hfill&\qquad ($2+M$ total),\hfill\cr}$$
in agreement with the earlier result for the Hodge numbers that we
deduced from the number of massless vector and hyper multiplets in the
$T^6/\IZ_2$ orientifold.


\newsec{Approximate versus exact M~theory lifts of orientifolds}

The standard relation \MmetricGeneral\ between the 11D metric and type
IIA supergravity background assumes that the 11d metric has an
isometry in the direction used in the dimensional reduction.  If this
is not the case, then the relation does not give the full 11D metric,
but rather its truncation to the lowest Fourier mode around the
M~theory circle.  To gain insight on the relation between the exact
Calabi-Yau duals of $T^6/\IZ_2$ and the leading order description
given in the previous section, it is helpful to first consider the
simpler case of an O6 plane and/or D6 branes in flat space.  The
description below draws heavily on Ref.~\Sen.

\subsec{The M~theory lift of a single D6 brane or O6 plane}

For the lift of a single D6 brane or O6 plane in flat space, the
M~theory background resulting from Eq.~\MmetricGeneral\ is the purely
geometrical background given by $\IR^{6,1}$ times a 4D space with
metric
\eqna\DOlift
$$ds^2_4 = Z^{-1}{R_{10}}^2\bigl(dx^{10}+A\bigr)^2 + Zds^2_{\IR^3},
\eqno\DOlift a$$ 
where
$$dA = R_{10}^{-1}*_3dZ,\quad\hbox{and}\quad 
Z = 1+{R_{10}\over8\pi^2}{Q\over\left|\vec x\right|}
\quad\hbox{with}\quad\vec x\in\IR^3.\eqno\DOlift {b,c}$$

For a D6 brane, we have $Q=1$ and this defines a smooth Taub-NUT
space.  Nothing is lost in the truncation to lowest Fourier mode since
the Taub-NUT space has a $U(1)$ isometry around the $x^{10}$ fiber.
The core of the D6 brane lifts to the point at which this fiber
shrinks to zero radius.

For an O6 plane, $Q=-4$.  Eq.~\DOlift{} together with the coordinate
identification $(\vec x, x^{10})\cong-(\vec x, x^{10})$ defines the
large radius approximation to an Atiyah-Hitchin space~\Atiyah.  It is
singular at small $\vec x$.  However, this just reflects the fact that
we have discarded all higher Fourier modes in the $x^{10}$ direction.
The complete Atiyah-Hitchin geometry is smooth.

The approximation becomes progressively worse as $\left|\vec x\right|$
is decreased, until at small but finite $\vec x$ the inverse warp
factor diverges and the M~theory circle decompactifies.  In type IIA
language, D0 branes become light near the O6 planes, where
$e^\phi\to\infty$.  In this region, the truncation of the low energy
effective field theory to the 10D type IIA supergravity multiplet is a
poor approximation.  It is necessary to include the RR charged fields
from the complete tower of D0 bound states to reliably describe the
local physics near the O6 planes.

The metric \DOlift{}\ also has an interpretation as the 1-loop moduli
space metric of a D2 brane probe~\ProbeBrane.  In the O6 case, this
metric is corrected by 3D instantons to the smooth Atiyah-Hitchin
metric.  However, this interpretation does not persist as an exact
quantitative correspondence in the case of primary interest in this
paper with only half as much supersymmetry~\DouglasGreene.  Therefore,
we will not pursue it here.\bigskip

To further set the stage for the interpretation of the approximate
Calabi-Yau metrics of \Sec{4}, we now review the $A_{M-1}$ and
approximate $D_M$ metrics resulting from leading order M~theory lift
of $M$ D6 branes (\Sec{5.2}) and $M$ D6 branes in the presence of an
O6 plane (\Sec{5.3}), respectively.  In either case, the metric is
again given by Eqs.~\DOlift{a,b}, however the expressions for the warp
factor differ.

\subsec{M~theory lift of $M$ D6 branes}

For the lift of $M$ D6 branes at positions $\vec x_I$ in the
transverse $\IR^3$, the warp factor is
\eqn\MDsix{Z = 1 +\sum_{I=1}^M Z^I,\quad\hbox{where}\quad
Z_I = {R_{10}\over8\pi^2}{1\over\left|\vec x-\vec x_I\right|}.}
Eq.~\DOlift{a,b}\ is then the metric for a multicentered Taub-NUT
space~\Hawking.  The $S^1$ fiber shrinks over each center $\vec x_I$
on the $\IR^3$ base.  In this geometry, we obtain 2-cycles with the
topology of spheres from the fibration over curves in $\IR^3$
connecting any pair of centers $x_I,x_J$.  In the notation of
Ref.~\Sen, if we let $S_{I,J}$ denote the holomology class of a sphere
formed from this pair of centers, then the set
$\{S_{1,2},S_{2,3},\ldots,S_{M-1,M}\}$ is a basis for $H_{2}$.  The
basis forms a ``chain of sausage links,'' in which neighboring spheres
intersect in 1 point: $S_{I-1,I}\cdot S_{I,I+1} = 1$.  Also,
${S_{I,I+1}}^2=-2$ since any two representatives of $S_{I,I+1}$
intersect in the points $x_I,x_{I+1}$.  The minus sign is due to
orientation.\foot{An alternative description of this minus sign is as
follows~\refs{\AspKthree,\AspTASI}.  A multicentered Taub-NUT space is
a Calabi-Yau 2-fold, so its first Chern class vanishes. One can then
show from the adjunction formula that the self-intersection number of
a genus $g$ holomorphic curve is $2g-2$.  The $g=0$ rational curves
are isolated, but formally have self-intersection $-2$ for agreement
with the $g>0$ formula.}  The intersection matrix is minus the Cartan
matrix of $SU(M)$, and the space describes the resolution of an
$A_{M-1}$ singularity.

There is also a Poincar\'e dual description.  There exists one $L_2$
harmonic form for each center $\vec x^I$~\TNHarmonic:
\eqna\FIDsix
$$\eqalignno{F^I 
&= \bigl(Z^I/Z\bigr)_{,m}\bigl(-dx^m\w(dx^{10}+A)
+\half ZR_{10}^{-1}\Vol_{\IR^3}{}^m{}_{np}\,dx^n\w dx^p\bigr)
& \FIDsix a\cr
&\equals^{\rm locally} d\Bigl(A^I - \bigl(Z^I/Z\bigr)
(dx^{10}+A)\Bigr). &\FIDsix b}$$
Here $A^I$ is defined by Eq.~\DOlift{b}\ with $Z$ replaced by $Z^I$.
The $F^I$ are anti-selfdual and have intersections
\eqn\FIDsixInt{\int F^I\w F^J = -\d^{IJ}.}
The harmonic forms $\omega^{I,J} = F^{I}-F^{J}$ have $A_{M-1}$
intersection matrix and are representatives of the cohomology classes
dual to the spheres $S_{I,J}$.  In discussing the Calabi-Yau duals of
$T^6/\IZ_2$, it will initially be more convenient to work with an
analogous cohomology description than with homology.

\subsec{The M~theory lift of $M$ D6 branes near an O6 plane}

For the lift of $M$ D6 branes plus one O6 plane at the origin, the
warp factor is

\eqna\MDOsix
$$Z = 1-{R_{10}\over8\pi^2}{4\over\left|\vec x\right|}
+\sum_{I=1}^M \bigl(Z^J+Z^{I'}\bigr),\eqno\MDOsix a$$
where
$$Z^I = {R_{10}\over8\pi^2}{1\over\left|\vec x-\vec x_I\right|}
\quad\hbox{and}\quad Z^{I'} = {R_{10}\over8\pi^2}
{1\over\left|\vec x+\vec x_I\right|},\eqno\MDOsix b$$
and there is a $\IZ_2$ coordinate identification $(\vec x,
x^{10})\cong-(\vec x, x^{10})$.  In this case, for each pair of
centers $\vec x_I,\vec x_J$, we obtain a sphere $S_{I,J}$ from the
$S^1$ fibration over a curve in $\IR^3$ connecting the two centers,
together with its $\IZ_2$ image.  In addition, for each center $\vec
x_I$ and image center $\vec x_{J'} = -\vec x_J$, we obtain a sphere
$S_{I,J'}$ from the $S^1$ fibration over a curve in $\IR^3$ connecting
$\vec x_I$ and $\vec x_{J'}$ together with its $\IZ_2$ image.  A basis
of $H_2$ is obtain from the classes $S_{I,I+1}$ for $I=1,\ldots,M-1$,
together with $S_{M-1,M'}$.  The intersection numbers in this basis
can be computed as in the previous case.  The intersection matrix is
minus the Cartan matrix of $SO(2M)$, and the metric is a large radius
approximation to that of a resolved $D_M$ singularity.

The metric is singular at sufficiently small $\left|\vec x\right|$.
However, we can choose representative cycles that avoid the bad
regions where $Z\le0$.  Therefore, we obtain correct topological data
(intersection numbers) from the singular leading order lift.  The full
lift smoothly excises the $Z\le0$ regions just as an Atiyah-Hitchin
space did for the lift of an O6 plane alone.

There is again a Poincar\'e dual description.  In this case, the
harmonic representative of the cohomology class dual to $S_{I,J}$ is
$\omega_{I,J} = (F^I-F^J) - (F^{I'}-F^{J'})$, and that of the
cohomology class dual to $S_{I,J'}$ is $\omega_{I,J'} = (F^I-F^{J'}) -
(F^{I'}-F^J)$.  Here, $F^I$ is given by Eq.~\FIDsix{}\ and $F^{I'}$ by
the same formula with $I$ replaced by $I'$.  Explicit calculation
shows that these cohomology classes give exactly a $D_{M}$
intersection matrix.  It is somewhat counterintuitive that we obtain
the correct result, since the domain of integration includes the
unreliable $Z\le0$ regions.  However, this result is expected since
intersection numbers must be the same for other cohomology
representatives that are supported away from the $Z\le0$ regions.


\newsec{(Co)homology, intersections, and fibration structure of
the CY$_3$ duals}

In \Sec{4}, we derived an approximate description of Calabi-Yau
manifolds $X_6$ dual to $T^6/\IZ_2$ with $\CN=2$ flux.  The approximate
metric \CYmetricM\ contains a function $Z'$, which determines the
overall scaling of the metric in the $x^{10}$ and the $x^{6,7,8}$
directions.  We showed that $Z'=1+(\nbar h/2s)\hat Z$, with $\hat Z =
\CO(1)$ as $h/s\to0$, so that by tuning $h/s$, we can make the unreliable
$Z'\le0$ regions smaller and smaller.  In the limit $h/s\to0$, the
leading order metric becomes arbitrarily good approximation at most
points.  Moreover, nothing special happens at the bad loci (fixed
points of $\CI_3$); these loci locally have the geometry of an
Atiyah-Hitchin space times $\IR^2$ in the exact description.
Therefore, we expect the homology and cohomology to be reliably
computable in the leading order description of the duality, just as in
the previous section.

\subsec{Cohomology and intersection numbers of the Calabi-Yau duals}

Expressing the K\"ahler form \JO\ in terms of $\hat Z$, we have
\eqn\JZhat{J = s\o_S + h\o_H,}
where
\eqna\omegaHSI
$$\eqalignno{\o_H &= \mbar \eta^4\w \eta^5 + \nbar
\hat Z dx^6\w dx^7 + \nbar dx^8\w \bigl(dx^{10} + A\bigr),
&\omegaHSI a\cr
\o_S &= 2dx^6\w dx^7. &\omegaHSI b}$$
As in \Sec{5}, we obtain one harmonic form for each D~brane before the
duality:
$$\o_I = F^I - F^{I'}, \eqno\omegaHSI c$$
where now
\eqn\CYFI{\eqalign{\strut F^I 
&= \bigl(Z'^I/Z'\bigr)_{,m}\bigl(-dx^m\w(dx^{10}+A)
+\half Z'R_{10}^{-1}\Vol_{\IR^2}{}^m{}_{np}\,dx^n\w dx^p\bigr)\cr
& \strut\qquad 
+ 2m \bigl(Z'^I/Z'\bigr)\bigl(\eta^4\w dx^7 - \eta^5\w dx^6\bigr).}}
Here, $Z'^I=G(x,x_I)$ in the notation of \Sec{3.2}.  If we define $A^I$
via $dA^I = R_{10}^{-1} *_3 dZ'^I$, then the $F_I$ again have the
local expression~\FIDsix{b}.  In the leading order description
\CYmetricM,  $\o_H,\o_S,\o_I$ form a basis for (the free part of)
$H^2(X_6,\IZ)$.  The corresponding intersection numbers are $A\cdot
B\cdot C = \half\int \o_A\w\o_B\w\o_c$, where the factor of 1/2 is due
to the $\IZ_2$ identification of coordinates \CYZtwo.  Letting $\CE_I$
denote the Poincar\'e dual of $\o_I$, one finds that
\eqn\Intersections{H^2\cdot S = 2\mbar\nbar,\quad H\cdot \CE_I\cdot
\CE_J = -\mbar \d_{IJ},\quad\hbox{others} = 0.}

To check that the harmonic forms $\o_H,\o_S,\o_I$ are correctly
normalized representatives of integer cohomology, we use the fact that
for a IIA Calabi-Yau compactification with no discrete torsion, the
continuous NS $B$-field moduli take values in $(2\pi)^2\a'
H^2(X_6,\IR)/H^2_{\rm free}(X_6,\IZ)$.  Since the periodicities of the
$B$-field moduli can be deduced from the periodicities of the dual
moduli in the $T^6/\IZ_2$ orientifold, the lattice $H^2(X_6,\IZ)$
follows.  

To verify the normalization of $\o_S$, we note that
$\bigl(c_{(4)\,4567}\bigr)_{T^6/\IZ_2} = (2\pi)^2\a'
\bigl(b_{(2)\,67}\bigr)_{\rm CY}$.  In this case, the dual periodicity
is $c_{(4)\,4567} \cong c_{(4)\,4567} + 2\cdot (2\pi)^4\a'^2$.  The
factor of 2 is related to the orientifold projection.\foot{In
Ref.~\FP, this factor of 2 follows from formulae in Secs.~III and IV
relating type $T^6/\IZ_2$ in the absence of flux to type I, and
subsequently to the heterotic string on $T^6$.  Another way to derive
it is to require that $\exp(i\oint c_{(4)})$ be single-valued for D3
instantons wrapping half-cycles on $T^6$ that descend to cycles on
$T^6/\CI_6$ in the $T^6/\IZ_2$ orientifold.}  Therefore, $2dx^6\w
dx^7$ generates a maximal 1D sublattice of $H^2_{\rm free}(X_6,\IZ)$
in the leading order description.  In the next section, we will
interpret the Calabi-Yau manifold $X_6$ as a fibration over $\IP^1
\cong T^2_{\{67\}}/\CI_2$.  The form $\o_S$ can then be interpreted as
the pullback of the generator of $H^2(T^2/\CI_2,\IZ)$ to $X_6$.

For $\o_H$, the check is slightly more complicated, since we need to
take into account moduli constraints.  In this case,
$\bigl(b_{(2)\,45}\bigr)_{\rm CY} =
(2\pi)^2\a'\bigl(c_{(0)}\bigr)_{T^6/\IZ_2}$ and
$\bigl(b_{(2)\,8\,10}\bigr)_{\rm CY} =
(2\pi)^2\a'\bigl(\Re(-1/\t_3)\bigr)_{T^6/\IZ_2}$.  The moduli $c_{(0)}$
and $\Re(-1/\t_3)$ have unit periodicity in the absence of flux;
however, the flux imposes the moduli constaint
\eqn\BConstraint{(n/m)\tdil = -1/\t_3\quad\Leftrightarrow\quad
n b_{(2)\,45} = mb_{(2)\,8\,10}.}
This moduli constraint implies the combined periodicity
\eqn\ReducedPer{\bigl(b_{(2)\,45},b_{(2)\,8\,10}\bigr)\cong
\bigl(b_{(2)\,45},b_{(2)\,8\,10}\bigr) +(2\pi)^2\a'(\mbar,\nbar),}
where, as above, we define $(\mbar,\nbar) = (m,n)/\gcd(m,n)$.
Comparing Eq.~\ReducedPer\ to \omegaHSI{b}, we see that $\o_H$ is
correctly normalized to generate a maximal 1D sublattice of $H^2_{\rm
free}(X_6,\IZ)$ in the leading order description.

For the $\o_I$, the check is cumbersome but straightforward.  We omit
the details here.  The harmonic forms \omegaHSI{c}\ are correctly
normalized as a consequence of the periodicities of the D3~brane
worldvolume scalars in the $T^6/\IZ_2$ orientifold, or equivalently,
the periodicities of the D6~brane worldvolume Wilson lines in the
O6/D6 dual.

The reader might wonder why there are only two explicit K\"ahler
moduli in Eq.~\JZhat, but $2+M$ harmonic forms.  The reason is that
K\"ahler deformations in the $\o_I$ directions can always be absorbed
into the warp factor.  Taking $J\to J+ t^I\o_I$ for small $t^I$
deforms $J$ to a nearby cohomology class whose harmonic representative
in the corresponding deformed metric takes exactly the same form as
\JZhat, but with slightly displaced sources entering into the
Poisson equation for $Z'$.

\subsec{Fibration structure}

The qualitative form of $X_6$ that arose in \Sec{4}\ from the duality
chain is as follows.  We first fiber $T^2_{\{4,5\}}$ over
$T^3_{\{678\}}$, then fiber $S^1_{\{10\}}$ over the resulting
geometry, and finally quotient by the $\IZ_2$ coordinate
identification \CYZtwo.  This means that in the leading order
description, the geometry is a fibration over $T^3_{\{678\}}/\CI_3$.
Since the latter is is nonorientable, this fibration is not very
useful.  However, we can instead view the geometry as a fibration over
$T^2_{\{67\}}/\CI_2\cong \IP^1$, where $\CI_2$ takes $(x^6,x^7)$ to
$-(x^6,x^7)$.\foot{The space $T^3/\CI_3$ is an $S^1$ fibration over
$T^2/\CI_2$ with singular fibers at the four $\CI_2$ fixed points.}
The field strengths of all connection 1-forms ($\CA^4,\CA^5,A,$ or
equivalently, $\CA^{z^1},\CA^{z^3}$) restrict trivially to the
subspace $(x^6,x^7) =$ constant.  Therefore, the generic fiber is the
product $S^1_{\{10\}}\times S^1_{\{8\}}\times T^2_{\{45\}} = T^4$,
with no twists.  We can trust this result since our approximate
description can be made valid to arbitrary accuracy away from the
$\CI_3$ fixed loci, and therefore suffices for describing the generic
fiber.  In fact, this $T^4$ has an additional structure that makes it
an abelian surface.

The K\"ahler form on the fiber is
\eqn\Jfiber{J_{\rm fib} = h\omega,}
where $h$ is the same K\"ahler modulus as above and $\o$ is the
2-form 
\eqn\ofiber{\o = \mbar dy^4\w dy^5 + \nbar dy^8\w dy^{10}}
on the fiber.  Here, $y^m$ are coordinates with unit periodicity
$y^m\cong y^m+1$, which may or may not be the same as the coordinates
$x^m$ restricted to a given point in the base, depending upon the
choice of gauge for the connections.  The possible values of
$(\mbar,\nbar)$ are $(1,1)$, $(1,2)$, $(1,3)$, $(1,4)$, together with
$\mbar\leftrightarrow\nbar$ interchanges.

In general, when $T^{2d}$ has a K\"ahler form proportional to
\eqn\Polarization{\o = \sum_{i=1}^d a_i dy^{2i-1}\w dy^{2i},\quad
\hbox{with}\quad a_i\mid a_{i+1},}
we can apply the Kodaira Embedding Theorem, which states that~\GandH:
{\it a compact complex manifold $X$ is an algebraic variety---i.e., is
embeddable in a projective space---if and only if it has a closed
$(1,1)$-form $\o$ whose cohomology class $[\o]$ is
rational.}\foot{An algebraic variety is locally the common
zero locus of set of homogeneous polynomials in a complex projective
space.}

In the case that $X=T^{2d}$, the variety is called an abelian variety
and the cohomology class $[\o]$ is called a polarization.  The
integers $a_i$ are the invariants of the polarization~\GandH.  In our
case, $d=2$, so the the torus is an abelian surface, and the
Calabi-Yau an abelian surface fibration over $\IP^1$.

\subsec{Homology}

We are now in a position to identify the homology classes Poincar\'e
dual to Eq.~\omegaHSI{a,b,c}.  The easiest to identify is the class
$S$ dual to $\o_S$, which is the class of the abelian surface fiber.

Since the fiber is an algebraic variety, it has at least a hyperplane
class of divisors (obtained by intersecting the variety with
codimension 1 hyperplanes of the projective space).  The class $H$
dual to $\o_H$ is the fibration of the hyperplane class of $T^4$ over
the $\IP^1$ base.

The remaining homology classes are very similar to those of
\Sec{5.3}.  The leading order description on the $\IZ_2$ cover is as
follows.  The class $\CE_I-\CE_J$ dual to $\o_I-\o_J$ is represented
by the fibration of $T^2_{\{45\}}\times S^1_{\{10\}}$ over a real
curve in $T^3_{\{678\}}$ connecting $x_I$ to $x_J$, plus its $\IZ_2$
image fibered over the curve connecting $x_{I'}$ to $x_{J'}$.  This
4-cycle is the analog of the 2-sphere $S_{I,J}$ of \Sec{5.3}: the
circle $S^1_{\{10\}}$ shrinks at the endpoints, but now $T^2_{\{45\}}$
(which does not shrink) is fibered as well.  Similarly, the class
$\CE_I+\CE_J$ dual to $\o_I+\o_J$ is represented by the
$T^2_{\{45\}}\times S^1_{\{10\}}$ fibration over a real curve in
$T^3_{\{678\}}$ connecting $x_I$ to $x_{J'}$, plus its $\IZ_2$ image.
This is the analog of $S_{I,J'}$ of \Sec{5.3}.  A complete basis of
$H_2(X_6,\IZ)$ in the leading order description is obtained from
$\CE_I-\CE_{I+1}$ for $I=1,\ldots,M-1$ and $\CE_{M-1}+\CE_M$.  On a
divisor of class $H$, the intersection matrix of the curves
corresponding to the divisors $\CE_I-\CE_{I+1}$ ($I=1,\ldots,M-1)$ and
$\CE_{M-1}+\CE_M$ is proportional to the $D_M$ Cartan matrix
(cf.~\Sec{5.3}\ and Eq.~\Intersections).  This reflects the fact that
there is enhanced $SO(2M)$ gauge symmetry when all $M$ D3 branes (and
$M$ image D3 branes) coincide with an O3 plane in the dual $T^6/\IZ_2$
orientifold.

Let us now give an alternative description of the last paragraph in
terms of the abelian surface fibration over $\IP^1$ and the
degenerations of this fibration.  At each point $(x^6,x^7) =
(x^6_I,x^7_I)$ on the $\IP^1$ = $T^2_{\{67\}}/\CI_2$ base, the $T^4$
degenerates to $T^2_{\{45\}}\times I_1$.  The $I_1$ factor is
$\bigl. T^2_{\{8,10\}} \bigr|_{(x^6_I,x^7_I)}$, which has a single
node at $x^8 = x^8_I$ where the $S^1_{\{10\}}$ circle degenerates to a
point (since $Z'^{-1}\to0$ in Eq.~\Mmetric).  Note that the same cycle
degenerates in each $I_1$.  The class $\CE_I-\CE_J$ is represented by
the fibration of $T^2_{\{45\}}$ times the degenerating cycle of
$T^2_{\{8,10\}}$ over a real curve in the $\IP^1$ base.  In the full
Calabi-Yau geometry, there are expected to be additional degenerations
from the Atiyah-Hitchin-like regions that excise the $Z'\le0$ regions
of the leading order description.  The class $\CE_I+\CE_J$ in the full
geometry is represented by a $T^2_{\{45\}}\times S^1$ fibration over a
real 1D locus in the $\IP^1$ that terminates at $x_I$ and $x_J$ as
well as at the locations of other fiber degenerations in the
Atiyah-Hitchin-like regions.  In the leading order description, this
locus is a real curve connecting $x_I$ to one of the four $\CI_2$
fixed points, plus a curve connecting $x_J$ to the same fixed point;
the $S^1$ that degenerates is $S^1_{\{10\}}$.  In the full Calabi-Yau
geometry, the description is the same in the reliable regions;
however, in the Atiyah-Hitchin-like regions, the 1D locus in the
$\IP^1$ base can become a more complicated junction of curves over
which different $S^1$ cycles of $T^2_{\{8,10\}}$ degenerate.

We can gain some intuition for the geometry in the Atiyah-Hitchin-like
regions as follows.\foot{The discussion in this paragraph is based on
Ref.~\MITjunction.  We refer the reader to the $D_M$ section of
Ref.~\MITjunction\ for a more complete exposition of the ideas
presented here.}  Consider an elliptic realization of the $D_M$ space
$X_4$ of \Sec{5.3}.  In the F~theory limit in which the area of the
fiber is scaled to zero, M~theory on $X_4$ becomes a type IIB
orientifold with one O7 plane and $M$ D7 branes, i.e., the IIB
background that is T-dual to the O6/D6 background of \Sec{5.3}.  In
the F~theory description, each D7 brane corresponds to an $I_1$
degeneration of the elliptic fibration in which a $(1,0)$ cycle of the
fiber degenerates.  The O7 plane resolves to a pair of $(p,q)$
seven-branes with $(p,q) = (1,1)$ and $(1,-1)$.\foot{This can be seen
in the Seiberg-Witten theory~\SW\ on a D3 brane probing an O7
plane~\BprobeB.  Classically, there is a point of enhanced $SU(2)$
symmetry where the D3 brane is coincident with the O7 plane.
Nonperturbatively, there is no enhanced $SU(2)$ point.  The point
resolves into a massless $(1,1)$ dyon point and a separate massless
$(1,-1)$ dyon point, corresponding to a D3 brane coincident with
either of the two types of $(p,q)$ seven-branes.  (The massless
hypermultiplets come from strings stretched between the D3 brane and
seven-brane.)}  Each $(p,q)$ seven-brane corresponds to an $I_1$
degeneration of the elliptic fibration of $X_4$ in which a $(p,q)$
cycle of the fiber degenerates.  The rational curves $S_{I,J}$ come
from the fibration of the $(1,0)$-cycle of the fiber over a real curve
in the base connecting two of the locations of $(1,0)$ $I_1$
degenerations.  The rational curves $S_{I,J'}$ come from the $S^1$
fibration over an H-shaped junction in the base, where $x_I$ and $x_J$
are the endpoints on the right edge of the letter H, and the locations
of the two $(p,q)$ degenerations are the endpoints on the left edge of
the letter H.  In the IIB description, this H-shaped junction is a
string junction with two $(1,0)$ strings on the right, a $(2,0)$
string in the middle, and $(1,1),(1,-1)$ strings on the left.  (The
fundamental strings terminate on D7 branes, and $(p,q)$ strings on
$(p,q)$ seven-branes.)  In the M~theory description, the junction
fattens out into a rational curve $S_{I,J'}$, where the $S^1$ that is
fibered over a $(p,q)$ segment of the junction in the base is the
$(p,q)$-cycle of the elliptic fibration.

We expect something similar to happen in the Calabi-Yau geometry, with
the 2-cycles of the previous paragraph replaced by 4-cycles due to the
extra two dimensions from $T^2_{\{45\}}$.  It is tempting to
conjecture that in each of the four Atiyah-Hitchin-like regions of the
Calabi-Yau, the $T^4$ fiber degenerates to $T^2_{\{45\}}\times I_1$ at
two points on the base $\IP^1$.  (Each pair of points would coalesce
in the leading order description to one of the four fixed point of
$\CI_2$ in $\IP^1 = T^2_{\{67\}}/\CI_2$.)  Intuitively, the
$(m,n)$-dependent global topology of the abelian surface fibration
should not effect the local geometry in these regions.  However, we do
not yet have a sufficiently complete understanding of the geometry in
the Atiyah-Hitchin-like regions to motivate this conjecture further.
We leave this task for the future.

\subsec{Check of fibration result}

As a check that the Calabi-Yau duals of $T^6/\IZ_2$ with $\CN=2$ flux
are abelian surface fibrations over $\IP^1$, recall the following
theorem due to Oguiso~\OguisoThm, which is reviewed in
Refs.~\refs{\AspUbiquity,\AspKthree}: {\it Let $X$ be a minimal
Calabi-Yau threefold.  Let $D$ be a nef divisor on $X$.  If the
numerical $D$-dimension of $D$ equals one then there is a fibration
$\Phi : X\to W$, where $W$ is $\IP^1$ and the generic fiber is either
a K3 surface or an abelian surface.}

The numerical $D$-dimension of a divisor is simply the largest integer
$n$ such that $D^n\ne0$.  From the intersection
numbers~\Intersections, we see that $S^2=0$, so $S$ is a divisor of
numerical $D$-dimension 1.  A divisor is nef (numerically effective)
if $D\cdot C\ge0$ for any algebraic curve $C$.  The space of nef
divisors arises as the closure of the space of ample divisors ($D\cdot
C>0$), which is important in realizing projective embeddings.
However, nefness is also a weaker analog of effectiveness.  A divisor
is effective if it is a formal linear combination of irreducible
analytic hypersurfaces with nonnegative coefficients~\GandH.

We now give a heuristic argument for why $S$ should be effective.  Due
to the pairing \CpxPairing, it is natural to attempt to define complex
coordinates via
$$z^1 = x^4+\t_2 x^5,\quad z^2 = x^6 +\t_2 x^7,\quad z^3 = x^8 + \t
x^{10}.$$
However, as a consequence Eq.~\CpxCurvatures, of these are not
holomorphic coordinates.  For example, in the gauge $\CA^{z^1} = 2nz^2
dx^8$, the globally defined $(1,0)$-form $\eta^{z^1}$ is
$$dz^1+ 2n z^2 dx^8 = d\bigl(z^1-2n x^8\bigr) +
2n\bigl(z^2+1\bigr)dx^8 = d\bigl(z^1-2n\t_2 x^8\bigr)
+ 2n \bigl(z^2+\t_2\bigr) dx^8,$$
from which we deduce the nonholomorphic identifications (i.e.,
transition functions)
$$ \bigl(z^1,z^2,z^3\bigr) \cong \bigl(z^1-2n\Re(z^3),z^2+1,z^3\bigr)
\cong \bigl(z^1-2n\t_2 \Re(z^3),z^2+\t_2,z^3\bigr).$$ 
Eq.~\CpxCurvatures\ implies that our tentative definitions of $z^1$
and $z^3$ need to be modified in order for the identifications to be
holomorphic.  However, it does not imply that the definition of $z^2$
needs to be modified.  Assuming that it is not modified, a divisor of
class $S$ is given by the holomorphic equation $z^2=\hbox{constant}$,
and is therefore an effective divisor.  It then follows from Oguiso's
theorem that $X_6$ a fibration over $\IP^1$.

To decide whether the generic fiber is K3 or an abelian surface, we
use the following expression~\refs{\AspUbiquity,\AspKthree} for the
Euler number of $S$, which follows from $S^2=0$ together with the
fact that both K3 and $T^4$ have vanishing first Chern class:
\eqn\EulerS{\chi(S) = \int_S c_2(S) = S\cdot c_2(X_6).}
If $\chi(S) = 0$ then the fiber is an abelian surface, and if $\chi(S)
= 24$ it is a K3 surface.

It is easy to evaluate the right hand side, since the quantity $S\cdot
c_2(X_6)$ is a familiar object that appears in the genus~1 topological
A-model amplitude~\BCOV\ on $X_6$,
\eqn\Fone{F_1\propto \sum_{\a=1}^{h^{1,1}(X_6)}
\bigl(D_\a\cdot c_2(X_6)\bigr) t^\a + \hbox{worldsheet instantons}.}
Here, the sum runs over a basis of $H^{1,1}$, with corresponding
K\"ahler moduli $t^\a$ and Poincar\'e dual divisors $D_\a$.  This
topological string amplitude enters into a curvature squared term
proportional to $\Re\int F_1\tr(R-R*)^2$ in the 4D low energy
effective action.  In the $T^6/\IZ_2$ orientifold, as in type I or the
heterotic string on $T^6$, $F_1$ is determined by Green-Schwarz
anomaly cancellation\foot{I thank A.~Dabholkar for emphasizing this
point to me.} to be
\eqn\FoneProptoDil{F_1\propto\tdil +\hbox{quantum corrections},}
where $\tdil = c_{(0)} +i/g_s$.  We will see in the next section that
$1/g_s$ of the $T^6/\IZ_2$ orientifold maps to the K\"ahler modulus
$h$ of the Calabi-Yau $X_6$, i.e., there is no $s$-dependence in the
first term of Eq.~\Fone.  Therefore, $\chi(S) = S\cdot c_2(X_6) =
0$, and the fiber is an abelian surface.


\newsec{Torsion 1-cycles and discrete isometries of $X_6$}

We saw in Eq.~\AxionKinetic\ that the $c_{(4)}$ scalars of the
$T^6/\IZ_2$ orientifold are axionically coupled to the $U(1)$ gauge
bosons $b_{(2)\,m\m}$ and $c_{(2)\,m\m}$, with charges given by the
3-form flux.  When the coupling is nonminimal and a component of
$c_{(4)}$ couples to $N b_{(2)\,m\m}$ or $N c_{(2)\,m\m}$, the
corresponding $U(1)$ gauge symmetry is only partially broken, leaving
a residual discrete gauge symmetry $\IZ_N$.  These discrete gauge
symmetries contain information about the torsion cycles and discrete
isometries of the Calabi-Yau duals, $X_6$.

Let us focus on the torsion 1-cycles and discrete isometries.  Suppose
that there is a $\IZ_N$ 1-cycle $\g$, i.e., a cycle such that $N\g$ is
trivial in $H_1(X_6,\IZ)$.  This means that a fundamental string can
be wrapped on $\g$ with winding number conserved modulo $N$.  The
$\IZ_N$ gauge field that couples to the winding charge is $\int_\g
dx^m b_{(2)\,m\m}$.  Similarly, suppose that there is a discrete
$\IZ_N$ isometry.  Then, in a coordinate system adapted to this
isometry, there is a vector $k^m$ and a discrete Kaluza-Klein gauge
field $V_\m$, such that the gauge transformations are $x^m\to x^m + \L
k^m$ and $V_\m \to V_\m - \partial_\m\L$, where $\L$ is a
$\IZ_N$-valued function of the coordinates $x^m$.

From the kinetic terms \AxionKinetic\ and flux \NtwoFlux, the discrete
gauge symmetries of the $T^6/\IZ_2$ orientifold and its Calabi-Yau
duals are
$$\vbox{\settabs\+& $T^6/\IZ_2$ field &\qquad Charge \qquad
&\quad IIA CY$_3$ field\quad &\quad Gauge symmetry\quad&\cr 
\+& $T^6/\IZ_2$ field &\qquad Charge \qquad
&\quad IIA CY$_3$ field\quad &\quad Gauge symmetry\quad&\cr 
\vskip5pt
\+&\hfill $c_{(2)4\m}$\hfill &\hfill$n$\hfill 
&\hfill $b_{(2)\,5\m}$\hfill &\hfill $\IZ_n$ (winding)\hfill&\cr
\+&\hfill $c_{(2)5\m}$\hfill &\hfill$n$\hfill
&\hfill $b_{(2)\,4\m}$\hfill &\hfill $\IZ_n$ (winding)\hfill&\cr
\+&\hfill $b_{(2)4\m}$\hfill &\hfill$m$\hfill
&\hfill $V^4{}_\m$\hfill &\hfill $\IZ_m$ (isometry)\hfill&\cr
\+&\hfill $b_{(2)5\m}$\hfill &\hfill$m$\hfill 
&\hfill $V^5{}_\m$\hfill &\hfill $\IZ_m$ (isometry)\hfill&\cr}$$
together with other discrete gauge symmetries that correspond to
higher dimensional torsion cycles.  We conclude that $H_1(X_6,\IZ) =
\IZ_n\times\IZ_n$ and that there is a $\IZ_m\times\IZ_m$ isometry in
the Calabi-Yau threefold $X_6(m,n)$.  This in turn tells us something
about the fundamental group of $X_6$, since $H_1(X_6,\IZ)$ is the
abelianization of $\pi_1(X_6)$, defined as $\pi_1(X_6)$ modulo its
commutator subgroup.  In particular, the fundamental group is
nontrivial for $n>1$.


\newsec{S-duality of $T^6/\IZ_2$ as T-duality of abelian surface fibers} 

In \Sec{2.2}\ we observed that interchange of the flux parameters
$(m,n)$ in the $T^6/\IZ_2$ orientifold can be interpreted as S-duality
followed by a $90^\circ$ rotation in the 89-directions.  On the other
hand, in \Sec{7}\ we found that there is a $\IZ_n\times\IZ_n$ winding
symmetry and $\IZ_m\times\IZ_m$ isometry in the Calabi-Yau dual
description.  Since T-duality interchanges winding and isometry (NS
B-field and metric), this suggests that $m\leftrightarrow n$
interchange can be interpreted as some type of T-duality of $X_6$.
Let us try to make this more precise.

In the $T^6/\IZ_2$ orientifold, S-duality acts on the string coupling
and string frame internal metric as\foot{Here, $\a' = \tilde\a'$ is a
common S-duality convention, but we will not need to specify a
convention for the purposes of this section.}
\eqn\Sduality{\eqalign{g_s &\to \tilde g_s = 1/g_s,\cr
g_{mn}/\a' &\to \tilde g_{mn}/\tilde\a' = g_{mn}/(g_s^2\a').}}
We can map these S-duality transformations to transformations of the
Calabi-Yau K\"ahler moduli $h$ and $s$.  For $h$, we have
\eqn\hversusg{{\mbar h\over(2\pi)^2\a'_{\rm CY}} =
{v'_1\over(2\pi)^2\a'_{\rm CY}} 
= {v'_1 R'_9\over(2\pi/M_{11})^3}
= {v'_1 R'_9\over g'_s (2\pi\sqrt{\a'})^3} = {1\over g_s},}
where primes denote quantities in the O6/D6 background of \Sec{3.2},
and the final $1/g_s$ on the right is in the $T^6/\IZ_2$ orientifold.
Therefore, the $m\leftrightarrow n$ interchange duality inverts the
K\"ahler modulus $h$ in string units:
\eqn\hInversion{{\mbar h\over(2\pi)^2\a'_{\rm CY}} \to
{\bar{\tilde m}\tilde h\over(2\pi)^2\tilde\a'_{\rm CY}} =
{(2\pi)^2\a'_{\rm CY}\over\mbar h},}
where $(\tilde m,\tilde n) = (n,m)$.  Since $h$ gives the size of the
abelian surface fiber (cf.~Eq.~\Jfiber), this makes precise the
geometrical interpretation of the $m\leftrightarrow n$ duality.  It is
T-duality of the entire abelian surface fiber and relates two
Calabi-Yau threefolds $X_6(m,n)$ and $X_6(n,m)$ of identical Hodge
numbers.

The modulus $s$ also gets rescaled in the $m\leftrightarrow n$ duality.
This rescaling is governed by the equation
\eqn\sRescaling{\biggl({2\tilde s\over(2\pi)^2\tilde\a'_{\rm CY}}\biggr)
\biggl({\bar{\tilde m}\tilde h\over(2\pi)^2\tilde\a'_{\rm CY}}\biggr) =
\biggl({2s\over(2\pi)^2\a'_{\rm CY}}\biggr)
\biggl({\mbar h\over(2\pi)^2\a'_{\rm CY}}\biggr).}


\newsec{Lessons from (non)candidates for explicit Calabi-Yau
constructions}

The most thoroughly understood class of Calabi-Yau threefolds is the
connected web of smooth Calabi-Yau threefolds that are hypersurfaces
in 4D toric varieties.  This class has been analyzed extensively by
Kreuzer and Skarke, who tabulated all 473,800,776 reflexive polyhedra
in four dimensions~\Kreuzer.  So, it is a logical first place to look
for the Calabi-Yau duals of $T^6/\IZ_2$.  We obtained a total of eight
such manifolds $X_6(m,n)$:
$$\matrix{M & \qquad & (h^{1,1},h^{2,1})\hfill
& \qquad &\hfill(m,n)\phantom{,}\hfill\cr
\noalign{\vskip-5pt}
\hbox to 15pt {\hrulefill} &&\hbox to 50pt {\hrulefill}
&&\hbox to 95pt {\hrulefill}\cr
\noalign{\vskip3pt}
\hfill0\hfill && (2,2)\hfill && (4,1),\quad (2,2),\quad (1,4)\hfill\cr
\hfill4\hfill && (6,6)\hfill && (3,1),\quad (1,3)\hfill\cr
\hfill8\hfill && (10,10)\hfill && (2,1),\quad (1,2)\hfill\cr
\hfill12\hfill && (14,14)\hfill && (1,1)\hfill}$$
To review, $M$ gives the number of divisors other than $S$ and $H$,
and $(m,n)$ determines the group $\IZ_n\times\IZ_n$ of torsion
1-cycles and discrete isometries $\IZ_m\times\IZ_m$.  The reduced pair
$(\mbar,\nbar) = (m,n)/\gcd(m,n)$ gives the polarization invariants of
the abelian surface fibers and determines the intersection numbers of
the threefold.

The hypersurface Calabi-Yau manifolds of Kreuzer and Skarke have
trivial fundamental group aside from 16 exceptional cases, so we can
at best expect to find the $n=1$ duals of $T^6/\IZ_2$ in this
class.\foot{The exceptional cases were described in Ref.~\KreuzerPALP\
and can be ruled out explicitly.  I thank M.~Kreuzer for pointing out
their existence and for subsequent email correspondence.}  Explicitly
searching for the above Hodge numbers in the database of
Ref.~\KreuzerWeb, we find only the pair $(14,14)$, which appears three
times.  So, there are three candidate hypersurface Calabi-Yau
threefolds.  Each of these threefolds can be shown to be fibered by
elliptic K3 surfaces in multiple ways~\KreuzerPALP.\foot{I am grateful
to B.~Florea for insights on these threefolds and for introducing me
to PALP~\KreuzerPALP.}  We do not have any reason to believe that our
(14,14) abelian surface fibered threefold is also an elliptic K3
fibration, nevertheless this a logical possibility.  On the other
hand, there seems be a folk theorem that abelian surface fibered
Calabi-Yau manifolds cannot be realized as toric
hypersurfaces~\GrassiKlemm, so we deem this possibility unlikely.

Looking back at Eq.~\CpxPairing, the appearance of the $T^2$ modular
parameters $\t_2$ and $\t$ suggests that we try to construct $X_6$ as
free quotient of the product
\eqn\threeE{A = E_{\t_2}\times E_{\t_2}\times E_\t}
of three elliptic curves, two of which are the same.  In fact, Oguiso
and Sakurai have considered an explicit construction of exactly this
type.\foot{This example can only be realized as a complete
intersection and not as a hypersurface~\AspComment, in agreement with
the folk theorem mentioned in the previous paragraph.}  In
Ref.~\OguisoCYQuotient, these authors prove that the only Calabi-Yau
threefolds that are given by quotients involving an abelian threefold
$A$ are of the form $X=A/G$, where $G$ is a finite automorphism group
acting freely on $A$, and $G=\IZ_2\times\IZ_2$ or $D_8$.  They refer
to $(A,G)$ in these two cases as Igusa's pair and Igusa's refined
pair, respectively.  In each case, the resulting Calabi-Yau threefold
is an abelian surface fibration over $\IP^1$.  Moreover, in the first
case $h^{1,1}=3$ and in the second case $h^{1,1}=2$.  So, the second
case is a candidate for our three Calabi-Yau duals with Hodge numbers
$(2,2)$.

However, this candidate can be ruled out for two reasons.  The simpler
reason is that $H_1(X,\IZ)$ is not $\IZ_n\times\IZ_n$.  From the
explicit construction provided for Igusa's refined pair (Example
(2.18) in Ref.~\OguisoCYQuotient), we can identify the generators of
$\pi_1(X)$ explicitly and then abelianize to obtain $H_1(X,\IZ) =
(\IZ_2)^6$.  On more general grounds, we can employ the Cartan-Leray
spectral sequence,\foot{I thank V.~Braun for explaining this and for
directing me to Chapter 8$^{\rm bis}$ of Ref.~\McCleary.} to obtain
the short exact sequence
\eqn\ShortExact{0\to H_1(G,\IZ)\to H_1(A/G,\IZ)\to H_1(A,\IZ)_G\to 0.}
Here, $H_1(G,\IZ)$ is the abelianization of $G$, which for $G=D_8$ is
$(\IZ_2)^2$.  The group $H^1(A,\IZ)_G$ is the $\IZ^6$ lattice that
defines the abelian threefold $A$, modulo identifications $x\cong gx$
for $g\in G$; in the case that $G=D_8$, one finds that this is
$(\IZ_2)^4$.  Therefore, $H_1(A/G,\IZ)= (\IZ_2)^2\ltimes (\IZ_2)^4$,
in agreement with the explicit computation.

The second reason is potentially more useful.  A Calabi-Yau manifold
$X$ is a quotient of an abelian threefold if and only if $c_2(X)=0$.
(See the discussion in Ref.~\OguisoCYQuotient\ and
Refs.~\refs{\Kobayashi,\SBW}\ contained therein.)  For the purposes of
\Sec{6.4}, we were content to show that $c_2(X_6)\cdot S = 0$.
However, with a little more work, we can compute $c_2(X_6)$
explicitly.  In the conventions of Ref.~\AspUbiquity, the precise
statement of Eqs.~\Fone\ and \FoneProptoDil\ is
\eqn\Fone{\eqalign{F_1 &= -{4\pi i\over12} \sum_{\a=1}
\bigl(D_\a\cdot c_2(X_6)\bigr) t^\a + \hbox{worldsheet instantons,}\cr
&= -{4\pi i\over12}(M+8)\tdil + \hbox{quantum corrections},}}
where the first line is the expression in the type IIA Calabi-Yau
compactification and the second line is the expression in the
$T^6/\IZ_2$ orientifold with $M$ D3 branes.  The term proportional to
$M$ is due to the D3 branes and the term proportional to $8$ is due to
the O3 planes~\DasguptaMukhi.  The normalization of the $t_\a$ is such
that the shift symmetry is $t_\a\cong t_\a+1$.  Using the result
\hversusg, we have
\eqn\tdilVersustH{\Im(\tdil) = \mbar h/\bigl((2\pi^2)\a'_{\rm
CY}\bigl) = \mbar t_H.}
Therefore,
\eqn\ctwoResult{H\cdot c_2(X_6) = \mbar(M+8)\quad\hbox{and}\quad
D_\a\cdot c_2(X_6) = 0\hbox{ otherwise.}} 
In particular, $c_2(X_6)\ne0$, which means that 
\eqn\NoGoResult{\hbox{$X_6$ is not the quotient of an abelian
threefold.}}

This result has important implications for worldsheet instantons
corrections in the type IIA compactification on $X_6$, since it is
expected that most (all?) Calabi-Yau manifolds that are not quotients
of abelian threefolds contain rational curves~\OguisoCYQuotient.
Moreover, if such instantons exist in the Calabi-Yau duals, then there
are corresponding worldsheet or D~instanton corrections to the
$T^6/\IZ_2$ orientifold with $\CN=2$ flux.  It would be very
interesting to identify explicit constructions of the Calabi-Yau duals
and settle this issue.  Either way we learn something new: the
nonexistence of rational curves would be a counterexample to a
conjecture that all Calabi-Yau manifolds other than quotients of
abelian threefolds have rational curves; their existence would imply
new corrections to the $T^6/\IZ_2$ orientifold that have not been
deduced by any other means.


\newsec{Conclusions and outlook}

Let us review the main results of this investigation.  We have seen
that the $T^6/\IZ_2$ orientifold with $\CN=2$ flux has standard type
IIA Calabi-Yau duals.  Depending upon the choice of flux parameters
$(m,n)$, we obtained eight possible dual Calabi-Yau threefolds
$X_6(m,n)$, with Hodge numbers
$$(h^{1,1},h^{2,1}) = (2,2)^3,\ (6,6)^2,\ (10,10)^2,\ {\rm and}
\ (14,14)^1.$$
Here, the superscripts indicate degeneracies which are further
distinguished by the integers $(m,n)$, satisfying
$$4mn= 16-M,\quad\hbox{where}\quad h^{1,1}=h^{2,1}=M+2.$$
The integers $(m,n)$ determine the discrete gauge symmetries of the
Calabi-Yau compactification,
$$\IZ_n\times\IZ_n\hbox{ winding,}\quad\hbox{and}\quad
\IZ_m\times\IZ_m\hbox{ isometry},$$
as well as the intersection numbers
$$H^2\cdot S = 2\mbar\nbar,\quad
H\cdot\CE_I\cdot\CE_J = -\mbar\d_{IJ},\quad\hbox{where}\quad
(\mbar,\nbar) = (m,n)/\gcd(m,n).$$
Finally, we have seen that the Calabi-Yau $X_6(m,n)$ is an abelian
surface ($T^4$) fibration over $\IP^1$ with polarization invariants
$(\mbar,\nbar)$.  We identified $m\leftrightarrow n$ interchange with
S-duality times a 90$^\circ$ rotation in the $T^6/\IZ_2$ orientifold,
and with T-duality of the entire abelian surface fiber in the dual
Calabi-Yau description.

We have not succeeded in providing explicit algebro-geometric
constructions of these Calabi-Yau manifolds.  However, by calculating
$c_2(X_6)$ we were able to demonstrate that they are not free
quotients of abelian threefolds ($T^6$).  This makes it very likely
that there exist rational curves in $X_6(m,n)$ and corresponding
instantons in both the Calabi-Yau and $T^6/\IZ_2$ descriptions.  If
this is so, then we have identified nonperturbative corrections in the
$T^6/\IZ_2$ orientifold that were not previously known.  If it is not
so, then we have identified Calabi-Yau manifolds that do not have
rational curves and that are not quotients of abelian threefolds,
which is also an interesting result.

By relating $\CN=2$, $T^6/\IZ_2$ vacua to standard type IIA Calabi-Yau
vacua, we have shown that the latter are part of a connected family in
the sense of~\Bubbles.  Without explicit constructions of the dual
Calabi-Yau manifolds, there remains the stronger question of whether
these vacua are connected to a large web through a common moduli
space.  Perturbative reasoning suggests that there are no extremal
transitions.  In the $T^6/\IZ_2$ orientifold, there are only $D3$
branes.  When two D3 branes meet, we do not perturbatively expect to
find a transition to a Higgs branch analogous to the transition that
takes place when a D3 brane dissolves into a D7 brane and become
worldvolume flux.  However, this reasoning can fail
nonperturbatively.\foot{I am indebted to S.~Kachru for emphasizing
this point.}  An explicit construction of the Calabi-Yau duals would
make it clear whether conifold transitions can occur.

For the future, we leave open the problem of explicitly constructing
the Calabi-Yau duals of the $T^6/\IZ_2$ based on the data provided
here.  We have seen that such duals always exist for $T^6/\IZ_2$ with
$\CN=2$ flux.  A similar statment should hold for the K3$\times
T^2/\IZ^2$ orientifold with $\CN=2$ flux.\foot{Indeed, this duality
has subsequently been studied by P.~Aspinwall~\AspFluxes.}  It
would be interesting to perform the required dualities, which in this
case amount to T-dualities in two of the K3 directions (i.e., mirror
symmetry) and one of the $T^2$ directions.  Finally, we would like to
know whether there also exist heterotic duals in the $\CN=2$ case, and
ultimately, to what extent flux compactifications have standard
fluxless duals in the phenomenologically relevant case of $\CN=1$
supersymmetry.


\bigskip\centerline{\bf Acknowledgements}\medskip

It is a pleasure to thank Bobby Acharya, Paul Aspinwall, Volker Braun,
Atish Dabholkar, Ron Donagi, Bogdan Florea, Andrew Frey, Antonella
Grassi, Shamit Kachru, and Martin Ro\v cek for helpful discussions and
useful references, as well as Per Berglund for correspondence during
early stages of the project.  In addition, I thank S.~Kachru,
P.~Tripathy, and S.~Trivedi for the enjoyable collaboration from which
this investigation was a continous outgrowth.  Finally, I am grateful
to the organizers of the Simons Workshop at SUNY Stony Brook and to
Stanford University for hospitality during the course of this work.
This work was supported in part by the DOE under contract
DE-FG03-92-ER40701.

\listrefs\bye